\documentclass[showpacs,preprintnumbers,amsmath,amssymb]{revtex4}

\usepackage{amsmath}
\usepackage{amssymb}
\usepackage{bm}
\usepackage{graphicx}
\usepackage{multirow}
\usepackage{textcomp}

\allowdisplaybreaks[1]

\begin{document}


\title{Dispersive derivation of the pion distribution amplitude}
\author{Hsiang-nan Li}
\affiliation{Institute of Physics, Academia Sinica,
Taipei, Taiwan 115, Republic of China}

\date{\today}

\begin{abstract}
We derive the dependence of the leading-twist pion light-cone distribution amplitude (LCDA) on
a parton momentum fraction $x$ by directly solving the dispersion relations for the  moments 
with inputs from the operator product expansion (OPE) of the corresponding correlation function. 
It is noticed that these dispersion relations must be organized into those for the Gegenbauer 
coefficients first in order to avoid the ill-posed problem appearing in the conversion from the moments 
to the Gegenbauer coefficients. Given the values of various condensates in the OPE, we find 
solutions for the pion LCDA, which are convergent and stable in the Gegenbauer expansion. Moreover, the 
solution from summing contributions up to 18 Gegenbauer polynomials is smooth, and can be well 
approximated by a function proportional to $x^p(1-x)^p$ with $p\approx 0.45$ at the scale $\mu=2$ GeV.
Turning off the condensates, we get the asymptotic form for the 
pion LCDA as expected. We then solve for the pion LCDA at a different scale $\mu=1.5$ GeV with the 
condensate inputs at this $\mu$, and demonstrate that the result is consistent with the one 
obtained by evolving the Gegenbauer coefficients from $\mu=2$ GeV to 1.5 GeV. That is, our formalism 
is compatible with the QCD evolution.  The strength of the above framework that goes beyond 
analyses limited to only the first few moments of a LCDA in conventional QCD sum rules is highlighted.   
The precision of our results can be improved systematically by including higher-order and higher-power
terms in the OPE.

\end{abstract}


\maketitle

%
%
%

\section{INTRODUCTION}

A light-cone distribution amplitude (LCDA), which describes the momentum fraction distribution 
of a parton in a hadron, is a nonperturbative fundamental input to the collinear factorization
for exclusive QCD processes with a large energy scale $Q$ \cite{LB79}. When the factorization 
holds, infrared divergences in radiative corrections to a process are absorbed into hadron LCDAs, 
and the remnant, being infrared finite, is calculable at the parton level in perturbation theory. 
A physical quantity is then factorized into a convolution of a hard kernel with hadron LCDAs in parton 
momentum fractions. The corresponding factorization theorem should be proved to all orders in the strong
coupling $\alpha_s$ and to certain power in $1/Q$. A LCDA, despite being nonperturbative, is
universal, {\it i.e.}, process independent. With this universality, a LCDA, determined by 
nonperturbative methods or extracted from experimental data, can be employed to make predictions for 
other exclusive processes involving the same hadron. Accurate knowledge of hadron LCDAs is thus crucial 
for enhancing the precision of predictions from the factorization theorem, and helps probing possible
new physics in exclusive processes, such as exclusive $B$ meson decays.

Tremendous efforts have been devoted to the derivation of hadrn LCDAs in the literature
\cite{Stefanis:2014nla}, all of which have their own strength and weakness. Lattice QCD
\cite{CK85,GK86,MS89,DGR,GHP,VMB,BDF,ABB,BCG,Bali:2017ude,Bali:2019dqc,Detmold:2021qln}, as a 
first-principle method, is powerful for extracting information of LCDAs. However, the 
computation is usually limited to the first few moments, which are not sufficient to reconstruct 
the complete dependence on a parton momentum fraction $x$. Though the formulation based on the 
quasicorrelation function \cite{Zhang:2017bzy} or Euclidean correlation functions in general 
\cite{Braun:2007wv,Bali:2017gfr,Bali:2018spj,Radyushkin:2017gjd,Radyushkin:2019owq} allows access 
to LCDAs in the entire $x$ range, the behavior near the endpoints of $x$ cannot be reliably predicted 
\cite{Zhang:2020gaj,Hua:2020gnw,Hua:2022kcm}. QCD sum rules, as one of the major analytical approaches 
to nonperturbative quantities, have been applied to studies of hadron LCDAs extensively 
\cite{CZ82,HWX,BF89,Ball:2004ye,Ball:2006wn,Ball:2007rt,BMS,Zhong:2014jla,Zhong:2021epq,Mikhailov:2021kty}. 
Besides the restriction to analyses of the first few moments, it is hard to estimate theoretical 
uncertainties arising from the naive parametrization of spectral densities based on the quark-hadron 
duality. Using the Dyson-Schwinger equations, one can calculate arbitrarily many moments of a LCDA 
in principle \cite{Chang:2013pq,Raya:2019vwr,Shi:2021nvg}. Nevertheless, results depend on the kernels 
adopted for solving the corresponding gap and Bethe-Salpeter equations \cite{Chang:2013pq}: the 
rainbow-ladder and dynamical-chiral-symmetry-breaking-improved kernels lead to the Gegenbauer coefficient 
$a_2^\pi = 0.23$ and $a_2^\pi = 0.15$ at the scale of 2 GeV, respectively, which are quite different. 
The Gegenbauer coefficients in meson LCDAs were determined through a global fit of  
perturbative QCD factorization formulas to measured branching ratios and direct $CP$ asymmetries in 
hadronic two-body $B$ meson decays recently \cite{Hua:2020usv}. However, this phenomenological 
approach relies on the precision of both the theoretical framework and experimental data 
sensitively.

We proposed to handle QCD sum rules \cite{SVZ} for nonpertrbative observables, such as the $\rho$ 
meson mass, as an inverse problem in \cite{Li:2020ejs}. The spectral density on the hadron side of a 
dispersion relation, established from a correlation function, is regarded as an unknown. 
The operator product expansion (OPE) for the correlation function on the quark side is calculated in the 
standard way without ambiguity like choices of the gap and Bethe-Salpeter kernels. The spectral density, 
including both resonance and continuum contributions, is then derived 
by solving the dispersion relation as an inverse problem with the OPE inputs. This formalism 
does not involve a continuum threshold, because the continuum contribution is supposed to be a smooth 
function, and may deviate from the perturbative one (the quark-hadron duality is likely to be broken).
It does not require a Borel transformation to suppress the continuum contribution and higher-power 
corrections to the OPE, and needs no discretionary stability criteria \cite{Coriano:1993yx,Coriano:1998ge} 
as usually postulated in conventional sum rules, once a dispersion relation is solved directly.
The long-existing concern on the rigorousness and predictive power of QCD sum rules 
\cite{Leinweber:1995fn,Gubler:2010cf} is then resolved. As an example, we demonstrated how to 
extract the masses and decay constants of the series of $\rho$ resonances from the dispersion 
relation obeyed by a two-current correlator \cite{Li:2020ejs}. We then developed an inverse matrix 
method to solve for scalar and pseudoscalar glueball masses from the corresponding dispersion relations 
\cite{Li:2021gsx}. This new viewpoint, based only on the analyticity of physical observables, has been 
extended to the explanation of the $D$ meson mixing parameters \cite{Li:2020xrz} and to the constraint 
on the hadronic vacuum polarization contribution to the muon anomalous magnetic moment \cite{Li:2020fiz}. 

We will apply the inverse matrix method developed in \cite{Li:2021gsx} to solve the dispersion relations 
for the moments of the leading-twist pion LCDA with the OPE inputs given in \cite{Zhong:2021epq}. 
The naive parametrization based on the quark-hadron duality and the discretionary requirement 
on the balance between perturbatve and condensate contributions to the OPE in conventional sum rules 
are not necessary. These improvements allow access to all moments of a LCDA as 
explained later. Since the buildup of an OPE is standard and reliable in the deep Euclidean region, the 
precision of our predictions can be enhanced by including higher-order and higher-power contributions
on the quark side systematically. We present the results of the moments $\langle\xi^2\rangle=0.2672$, 
$\langle\xi^4\rangle=0.1333$, $\langle\xi^6\rangle=0.0871$, $\langle\xi^8\rangle=0.0658$,
$\langle\xi^{10}\rangle=0.0546$, $\langle\xi^{12}\rangle=0.0480$,..., at the scale of 2 GeV, which
exhibit good convergence. The value of the second moment $\langle\xi^2\rangle$ is consistent with those 
from sum rules \cite{Zhong:2021epq} and lattice QCD \cite{Detmold:2021qln}, 
but lower than from the global fit to the data of hadronic two-body $B$ meson decays 
\cite{Hua:2020usv}. Note that the global analysis in \cite{Hua:2020usv} is based on the leading-order
factorization formulas, so its results are expected to be modified by subleading effects. In 
principle, we can evaluate the moments higher than $\langle\xi^{12}\rangle$ in our formalism, but will 
not continue due to the reason below. 

To construct the $x$ dependence of the pion LCDA from the obtained moments, one converts
the moments to the Gegenbauer coefficients, and then adds up the contributions from the Gegenbauer 
polynomials multiplied by the corresponding Gegenbauer coefficients. However, this conversion
involves a matrix, whose elements grow dramatically fast with the number of moments, i.e., the 
dimension of the matrix. Tiny errors in the obtained moments are then enlarged significantly during 
the conversion, such that the Gegenbauer coefficients go out of control unavoidably, as noticed in
\cite{Zhong:2021epq}. It turns out 
that the resultant pion LCDA becomes unstable in the Gegenbauer expansion, and reveals violent 
oscillations in the $x$ distribution. To overcome the difficulty, which originates from an ill-posed
nature, we first organize the dispersion relations into those for the Gegenbauer coefficients using 
the aforementioned matrix. A regulator is introduced into the matrix to stabilize the solutions for the 
Gegenbauer coefficients. It will be shown that these solutions are convergent with
the number of Gegenbauer coefficients, and insensitive to the regularization. Moreover, the shape of
the pion LCDA is smooth after summing the contributions up to 18 Gegenbauer polynomials,
which can be well described by a function proportional to $x^p(1-x)^p$ with $p=0.45\pm 0.02$ 
at the scale $\mu=2$ GeV. This shape is close to the one from the 
dynamical-chiral-symmetry-breaking-improved kernel for the Dyson-Schwinger equations 
\cite{Chang:2013pq,Roberts:2021nhw}, and a bit narrower than from the recent lattice QCD calculation 
based on the quasi-light-front correlation function \cite{Hua:2022kcm}. 

We stress that the pion LCDA can be derived at any scale $\mu$ in our framework by choosing 
the condensate values  in the OPE at a designated $\mu$. On the other hand, each of the 
Gegenbauer coefficients in a LCDA follows the well-known QCD evolution governed by a specific 
anomalous dimension. Hence, it is worth investigating how well our method is compatible 
with the QCD evolution. To do it, we solve for the pion LCDA at another scale $\mu$, say 
$\mu=1.5$ GeV, directly from the dispersion relations with the inputs at this $\mu$, and also 
get the pion LCDA at $\mu=1.5$ GeV through the evolution of the Gegenbauer coefficients obtained 
at $\mu=2$ GeV. It will be observed that these two results agree with 
each other within theoretical uncertainties. We speculate that their minor distinction is attributed
to the incomplete $\mu$ dependence of the currently available OPE, and that the inclusion 
of higher-order contributions into the OPE will improve the agreement. The above investigation 
supports the consistency of our formalism.

The rest of the paper is organized as follows. In Sec.~II we establish the dispersion relations 
for the moments and for the Gegenbauer coefficients of the pion LCDA, and highlight their advantages 
over conventional sum rules. The inverse matrix method to solve the dispersion relations is also 
elaborated. In Sec.~III we first validate our approach by solving for a sample LCDA from the 
dispersion relations with the inputs of the mock data, which are generated by the sample LCDA. The 
successful reproduction of the sample LCDA from the mock data encourages the application to 
exploring the realistic pion LCDA. The dispersion relations are then solved with the condensate 
inputs given in the literature, and the moments and the Gegenbauer coefficients of the pion LCDA are 
determined. In particular, we get the asymptotic form for the pion LCDA in the absence of the 
condensates. The theoretical uncertainty mainly comes from the dimension-six condensates, which cause 
about 10\% errors to our results. The stability and reliability of the obtained LCDA, and the 
compatibility of our formalism with the QCD evolution are also justified. Section IV contains the 
conclusion and outlook. The reformulation of the dispersion relations to facilitate the
inverse matrix method is detailed in the Appendix.

\section{FORMALISM}

The sum rules for the leading-twist pion LCDA were deduced from a two-point correlation function in the 
framework of the background field theory \cite{HWX,Zhong:2014jla}, and refined, together with the numerical analysis, 
in \cite{Zhong:2021epq}. To illustrate their restrictive application to the derivation of the moments, 
we quote the explicit expression
\begin{eqnarray}
\frac{f_\pi^2\langle\xi^n\rangle \langle\xi^0\rangle }{M^2e^{m_\pi^2/M^2}}
&=&\frac{3}{4\pi^2(n+1)(n+3)}\left(1-e^{-s_\pi/M^2}\right)
+\frac{m_u\langle\bar uu\rangle+m_d\langle\bar dd\rangle}{M^4}
+\frac{1}{12\pi}\frac{\langle \alpha_sG^2\rangle}{M^4}\nonumber\\
& &-\frac{8n+1}{18}\frac{m_u\langle g_s\bar u\sigma TGu\rangle+m_d\langle g_s\bar d\sigma TGd\rangle}{M^6}
+\frac{2(2n+1)}{81}\frac{\langle g_s\bar uu\rangle^2+\langle g_s\bar dd\rangle^2}{M^6}
-\frac{n\theta(n-2)}{48\pi^2}\frac{\langle g_s^3fG^3\rangle}{M^6}\nonumber\\
& &+\frac{1}{486\pi^2}\left\{\left[2(51n+25)-2n\theta(n-2)\right]\ln\frac{M^2}{\mu^2}+ C_n\right\}
\frac{\sum_{u,d,s}\langle g_s^2\bar\psi\psi\rangle^2}{M^6},\label{qsr}
\end{eqnarray}
with the coefficient
\begin{eqnarray}
C_n=3(17n+35)+\theta(n-2)\left\{\frac{49n^2+100n+56}{n}-25(2n+1)
\left[\psi\left(\frac{n+1}{2}\right)-\psi\left(\frac{n}{2}\right)+\ln 4\right]\right\}.\label{bn}
\end{eqnarray}
The left-hand side of Eq.~(\ref{qsr}) arises form the pion pole contribution, where $f_\pi$ ($m_\pi$) is the 
pion decay constant (mass), $M$ is the Borel mass, and the $n$th moment 
$\langle\xi^n\rangle$ with $\xi=2x-1$ is defined via the pion LCDA $\phi_\pi(x)$ by
\begin{eqnarray}
\langle\xi^n\rangle\equiv\int_0^1 dx(2x-1)^n\phi_\pi(x).\label{mel}
\end{eqnarray}
The right-hand side is a result of the OPE for the two-point correlation function, where the first term 
denotes the perturbative contribution with the threshold $s_\pi$ being introduced through the 
parametrization of the continuum contribution based on the quark-hadron duality, the others 
are the power corrections proportional to various quark and gluon condensates with the light quark masses 
$m_u$ and $m_d$, and the strong coupling $g_s$. The $\theta$-function $\theta(n-2)$ takes the value of 
unity as $n\ge 2$, and $\psi$ in Eq.~(\ref{bn}) represents the polygamma function. As to the notations in 
the condensates, $G$ ($\psi=u$, $d$, or $s$) is the gluon (quark) field, $\sigma$ ($T$) stands for a 
Dirac-gamma (color) matrix, and $f$ abbreviates a SU(3) structure constant.

It is immediately noticed that the perturbative (condensate) contribution decreases (increases) with the integer 
$n$. Namely, the convergence of the OPE deteriorates with $n$, such that
the calculation of higher moments is not reliable. One may choose a large $M$ to suppress the
power corrections. At the same time, the value of $s_\pi$ should not exceed the mass squared of the next
excited state, otherwise the single-pole parametrization, which leads to Eq.~(\ref{qsr}), fails.
The factor $1-e^{-s_\pi/M^2}$, diminishing with $M$ for a finite $s_\pi$, then lowers the
perturbative contribution. Therefore, lifting $M$ does not improve the convergence of 
the OPE effectively, and it is why sum-rule analyses are limited to the first few moments. 
A discretionary criterion was imposed on the power corrections with the dimension-six contribution to a moment 
being smaller than 5\% of the total one in \cite{Zhong:2021epq}. Combining the other criteria with the 
perturbative contributions being above $30\%, 35\%, 40\%, 40\%, 40\%$ for $n = 2, 4, 6, 8, 10$, respectively, 
the authors fixed the Borel windows in $M$, within which the corresponding moments were computed \cite{Zhong:2021epq}. 
It is obvious that the above prescriptions induce theoretical uncertainties, which are not easy to estimate 
rigorously. Moreover, extracting the Gegenbauer coefficients from the moments, i.e., the conversion from
the Mellin space to the $x$ space is also an ill posed problem, and suffers 
large uncertainties, especially when many moments are involved. That is, getting more
moments will not guarantee the uncovering of the correct $x$ dependence of a LCDA, if the precision of the 
moments is not high enough. This is another major obstacle for applying conventional sum rules
to studies of LCDAs. As shown in Sec.~III, the difficulty and ambiguity encountered in conventional 
sum rules can be avoided in our formalism.

Note that the sum rules in Eq.~(\ref{qsr}) have to be reformulated in order to be employed in 
our framework. The last line contains the power-suppressed logarithm $\ln M^2/M^6$,
which is proportional to $\ln(-q^2)/(q^2)^3$ before the Borel transformation, $q$ being the momentum 
flowing through the correlation function. To implement the inverse matrix method in \cite{Li:2021gsx}, 
both sides of a dispersion relation must be expanded into power series in $1/q^2$ (without the 
logarithm $\ln(-q^2)$), because the matrix equation to be solved is constructed by equating the 
coefficients of each power in $1/q^2$ on the two sides. If the logarithm exists,
such equating of the coefficients will not be legitimate. Hence, the power-suppressed 
logarithm needs to be handled in a nontrivial way to be explained in the next subsection.
We mention that choosing the scale $\mu^2=q^2$ does not resolve this issue, which just moves the 
logarithmic dependence to the strong coupling constant $\alpha_s(q^2)$ in the OPE.

\subsection{Reformulation of the Dispersion Relation}

We consider the same correlation function $I_n(q^2)$ as in \cite{Zhong:2021epq}, and write the identity
\begin{eqnarray}
I_n(q^2)=\lim_{r\to 0}\frac{1}{\pi}\int_r^R ds\frac{{\rm Im}I_n(s)}{s-q^2}
+\frac{1}{2\pi i}\int_{C_R} ds\frac{I^{\rm pert}_n(s)}{s-q^2},\label{3}
\end{eqnarray}
by following the procedure in \cite{Li:2021gsx}.
The contour on the right-hand side of Eq.~(\ref{3}) consists of two pieces of horizontal paths 
above and below the branch cut along the positive real axis on the complex $s$ plane and a counterclockwise 
circle $C_R$ of large radius $R$ \cite{Li:2021gsx}. The spectral density ${\rm Im}I_n(s)/\pi$
is the unknown function, which collects nonperturbative contributions from the low $s$ region, and the 
perturbative piece $I^{\rm pert}_n(s)$ is chosen as an appropriate
expression for $I_n(s)$ in the region far away from physical poles.
The idea of reformulating the dispersion relation is to absorb the power-suppressed logarithmic term in the 
OPE, together with the perturbative contribution, into the contour integration of $I^{\rm pert}_n(s)$, 
\begin{eqnarray}
I_n^{\rm OPE}(q^2)&=&\frac{1}{2\pi i}\oint ds\frac{I_n^{\rm pert}(s)}{s-q^2}+I_n^{\rm cond}(q^2),\label{10}\\
I_n^{\rm cond}(q^2)&=&\frac{m_u\langle\bar uu\rangle+m_d\langle\bar dd\rangle}{(q^2)^2}
+\frac{1}{12\pi}\frac{\langle \alpha_sG^2\rangle}{(q^2)^2}
+\frac{8n+1}{9}\frac{m_u\langle g_s\bar u\sigma TGu\rangle+m_d\langle g_s\bar d\sigma TGd\rangle}{(q^2)^3}\nonumber\\
& &+\frac{n\theta(n-2)}{24\pi^2}\frac{\langle g_s^3fG^3\rangle}{(q^2)^3}
-\frac{4(2n+1)}{81}\frac{\langle g_s\bar uu\rangle^2+\langle g_s\bar dd\rangle^2}{(q^2)^3}
-\frac{C_n}{243\pi^2}\frac{\sum_{u,d,s}\langle g_s^2\bar\psi\psi\rangle^2}{(q^2)^3},\label{con6}
\end{eqnarray}
where the OPE is given in the $q^2$, instead of $M^2$, space.
The contour for the first term on the right-hand side of Eq.~(\ref{10}) consists of a small 
clockwise circle $C_r$ of radius $r$ around the origin, in addition to those in Eq.~(\ref{3}).
Compared to the OPE in Eq.~(\ref{qsr}), the power-suppressed logarithm is absent in
$I_n^{\rm cond}(q^2)$, which has been shifted into the first term.

The contour integration of the power-suppressed logarithm $L(q^2)=\ln(-q^2/\mu^2)/(q^2)^3$ yields,
as derived in Appendix A, 
\begin{eqnarray}
L(q^2)&=&\lim_{r\to 0}\left[\frac{1}{(q^2)^3}
-\frac{1}{(q^2)^2}\frac{d}{dr}+\frac{1}{2q^2}\frac{d^2}{dr^2}\right][-r^3L(-r)]\nonumber\\
& &+\lim_{r\to 0}\frac{1}{\pi}\int_r^R ds\frac{{\rm Im}L(s)}{s-q^2}
+\frac{1}{2\pi i}\int_{C_R} ds\frac{L(s)}{s-q^2},\label{i3}
\end{eqnarray} 
where the first (second, third) term on the right-hand side comes from the integral along the circle $C_r$
(the horizontal paths above and below the branch cut, the circle $C_R$). It is easy to see that the first 
two terms diverge as $r\to 0$, but the divergences cancel between them. Namely, the first term serves as 
an infrared regulator of the second term. With Eq.~(\ref{i3}),
the contour integral on the right-hand side of Eq.~(\ref{10}) becomes
\begin{eqnarray}
\frac{1}{2\pi i}\oint ds\frac{I_n^{\rm pert}(s)}{s-q^2}=\lim_{r\to 0} P_n(q^2,r)+
\lim_{r\to 0}\frac{1}{\pi}\int_{r}^R ds\frac{{\rm Im}I_n^{\rm pert}(s)}{s-q^2}
+\frac{1}{2\pi i}\int_{C_R} ds\frac{I_n^{\rm pert}(s)}{s-q^2},
\end{eqnarray} 
with the infrared regulator
\begin{eqnarray}
P_n(q^2,r)&=&-\frac{1}{243\pi^2(q^2)^3}\left[\ln\frac{r}{\mu^2}-\frac{q^2}{r}-\frac{(q^2)^2}{2r^2}\right]
\left[2(51n+25)-2n\theta(n-2)\right]
\sum_{u,d,s}\langle g_s^2\bar\psi\psi\rangle^2,\label{pn}
\end{eqnarray}
and 
\begin{eqnarray}
\frac{1}{\pi}{\rm Im}I_n^{\rm pert}(s)&=&\frac{1}{\pi} I_n^{\rm pert(1)}
+\frac{1}{\pi}{\rm Im} I_n^{\rm pert(2)}(s),\nonumber\\
\frac{1}{\pi}{\rm Im}I_n^{\rm pert(1)}&=&\frac{3}{4\pi^2(n+1)(n+3)},\nonumber\\
\frac{1}{\pi}{\rm Im}I_n^{\rm pert(2)}(s)&=&
\frac{1}{243\pi^2s^3}\left[2(51n+25)-2n\theta(n-2)\right]
\sum_{u,d,s}\langle g_s^2\bar\psi\psi\rangle^2.\label{ope2}
\end{eqnarray}
The second (last) line of the above expressions arises from the perturbative contribution 
(power-suppressed logarithmic term) in the OPE, so the perturbative piece $I^{\rm pert}_n(s)$
contains the quark condensate actually. 
The equality of Eqs.~(\ref{3}) and (\ref{10}) then establishes the modified dispersion relation
\begin{eqnarray}
\frac{1}{\pi}\int_{0}^R ds\frac{{\rm Im}I_n(s)}{s-q^2}=\lim_{r\to 0} P_n(q^2,r)+
\lim_{r\to 0}\frac{1}{\pi}\int_{r}^R ds\frac{{\rm Im}I_n^{\rm pert(1)}+{\rm Im} I_n^{\rm pert(2)}(s)}{s-q^2}+
I_n^{\rm cond}(q^2),\label{di4}
\end{eqnarray}
where the integrals along the contour $C_R$ have cancelled from 
both sides. Equation~(\ref{di4}) realizes the goal that all its terms can be expanded into power series 
in $1/q^2$, as will be demonstrated in the next subsection.

The width of a pion, being smaller than 10 eV, justifies the narrow-width
approximation for parametrizing the resonance contribution to the spectral density.
We thus decompose the spectral density on the hadron side into a pole contribution plus a 
continuum contribution,
\begin{eqnarray}
\frac{1}{\pi}{\rm Im}I_n(s)=f_\pi^2\langle\xi^n\rangle\langle\xi^0\rangle\delta(s-m_\pi^2)+\rho_n(s),
\label{pap}
\end{eqnarray}
where the unknown moment $\langle\xi^n\rangle$ and continuum function $\rho_n(s)$ 
will be obtained by solving the dispersion relation directly. The function $\rho_n(s)$ 
vanishes at the physical threshold $s\to 0$ like $\rho_n(s)\sim s$ \cite{Veliev:2010di}, and the only 
requirement on its behavior is that it approaches to the perturbative result as $s\to\infty$,
\begin{eqnarray}
\lim_{s\to\infty}\rho_n(s)\to\frac{3}{4\pi^2(n+1)(n+3)}.\label{17}
\end{eqnarray}
That is, the quark-hadron duality for the unknown continuum contribution is not assumed 
at any finite $s$ in the above construct.

Following \cite{Li:2021gsx}, we introduce a subtracted continuum function $\Delta\rho_n$, which is related to the 
original $\rho_n$ via
\begin{eqnarray}
\Delta\rho_n(s,\Lambda)=\rho_n(s)-\frac{1}{\pi}{\rm Im}I_n^{\rm pert(1)}[1-\exp(-s/\Lambda)]
-\frac{1}{\pi}{\rm Im}I_n^{\rm pert(2)}(s)[1-\exp(-s^4/\Lambda^4)].\label{sub}
\end{eqnarray}
The scale $\Lambda$ characterizes the order of $s$, at which $\rho_n(s)$ transits to the perturbative 
expression in Eq.~(\ref{17}). The smooth function $1-\exp(-s/\Lambda)$ ($1-\exp(-s^4/\Lambda^4)$)
vanishes like $s$ ($s^4$); namely, the subtraction terms decrease like $s$ in the $s\to 0$ limit
with ${\rm Im}I_n^{\rm pert(1)}$ being a constant and  ${\rm Im}I_n^{\rm pert(2)}(s)\sim 1/s^3$ as indicated 
in Eq.~(\ref{ope2}), such that the low-energy behavior of $\rho_n(s)\sim s$ is not altered. 
These smooth functions approach to unity at large $s\gg\Lambda$, rendering the dispersive 
integration of the subtracted continuum function converge. That is, they play the role of 
an ultraviolet regulator for a dispersive integral mentioned in 
\cite{Forkel:2003mk}. We have tested choices of other smooth functions, such as $1-\exp(-s^2/\Lambda^2)$ 
for ${\rm Im}I_n^{\rm pert(1)}$ or $1-\exp(-s^5/\Lambda^5)$ for ${\rm Im}I_n^{\rm pert(2)}(s)$, which 
diminishes more rapidly as $s\to 0$, and confirmed that our solutions for the moments remain untouched basically: 
the former (latter) replacement leads to only $O(1\%)$ ($O(0.1\%)$) reduction of the outcomes for the second
moment. Since $\Delta\rho_n(s,\Lambda)$ decreases quickly as $s>\Lambda$, the radius $R$ in Eq.~(\ref{di4}) 
can be pushed toward infinity, when the dispersion relation is formulated in terms of the subtracted 
continuum function,
\begin{eqnarray}
\frac{f_\pi^2\langle\xi^n\rangle\langle\xi^0\rangle}{m_\pi^2-q^2}
+\int_{0}^\infty ds\frac{\Delta\rho_n(s,\Lambda)}{s-q^2}
&=&\lim_{r\to 0} P_n(q^2,r)+I_n^{\rm cond}(q^2)\nonumber\\
& &+\lim_{r\to 0}\frac{1}{\pi}\int_{r}^\infty ds
\frac{{\rm Im}I_n^{\rm pert(1)}\exp(-s/\Lambda)
+{\rm Im} I_n^{\rm pert(2)}(s)\exp(-s^4/\Lambda^4)}{s-q^2},\label{r20}
\end{eqnarray}
where Eq.~(\ref{pap}) has been inserted.
The results for the moments $\langle\xi^n\rangle$ should be insensitive to the variation of the
transition scale $\Lambda$, which is introduced through the ultraviolet regulation of the
spectral density. The numerical analysis to be performed in the next section does verify this 
stability of $\langle\xi^n\rangle$.

\subsection{Inverse Matrix Method for Moments}

We illustrate how to solve the dispersion relation in Eq.~(\ref{r20}) as an inverse problem in the inverse 
matrix method \cite{Li:2021gsx}. The subtracted continuum function $\Delta\rho_n(s,\Lambda)$ is a 
dimensionless quantity, so it can be cast into the form $\Delta\rho_n(s/\Lambda)$. Certainly,  $\Delta\rho_n$ 
may depend on other constant scales, such as masses of excited states, which appear as constant 
ratios over a given $\Lambda$, instead of variables like $s/\Lambda$. 
Equation~(\ref{r20}) then reduces, under the variable changes $q^2=x\Lambda$,
$s=y\Lambda$ and $r=\epsilon\Lambda$, to
\begin{eqnarray}
\frac{r_f\langle\xi^n\rangle\langle\xi^0\rangle}{x -r_m}+\int_{0}^\infty dy\frac{\Delta\rho_n(y)}{x-y}
&=&-\lim_{\epsilon\to 0} P_n(x\Lambda,\epsilon\Lambda)-I_n^{\rm cond}(x\Lambda)\nonumber\\
& &+\lim_{\epsilon\to 0}\frac{1}{\pi}\int_{\epsilon}^\infty dy
\frac{{\rm Im}I_n^{\rm pert(1)}e^{-y}+{\rm Im} I_n^{\rm pert(2)}(y\Lambda)e^{-y^4}}{x-y},
\label{r21}
\end{eqnarray}
with the constant ratios $r_f=f_\pi^2/\Lambda$ and $r_m=m_\pi^2/\Lambda$. 
It is found that the transition scale $\Lambda$ in the subtracted continuum function has moved into 
the condensate terms on the right-hand side to make them dimensionless.

To express the involved quantities into matrix forms, we rewrite the index $n$ as $2n-2$, so that
$n$ runs over $1,2,3,\cdots$, instead of $0,2,4,\cdots$. Considering the boundary condition of 
$\Delta\rho_{2n-2}(y)\sim y$ at $y\to 0$ in Eq.~(\ref{sub}), we expand $\Delta\rho_{2n-2}(y)$ in terms 
of the generalized Laguerre polynomials $L_j^{(\alpha)}(y)$ with the parameter $\alpha=1$,
\begin{eqnarray}
\Delta\rho_{2n-2}(y)=\sum_{j=1}^N a_{jn}y^\alpha e^{-y}L_{j-1}^{(\alpha)}(y),\label{r0}
\end{eqnarray}
up to degree $N-1$, where the unknown coefficients $a_{jn}$ will be obtained in the inverse matrix method. 
The generalized Laguerre polynomials satisfy the orthogonality
\begin{eqnarray}
\int_0^\infty y^\alpha e^{-y}L_i^{(\alpha)}(y)L_j^{(\alpha)}(y)dy=\frac{\Gamma(i+\alpha+1)}{i!}\delta_{ij}.
\label{or0}
\end{eqnarray}  
As explained in \cite{Li:2021gsx}, the set of generalized Laguerre polynomials is the only choice of the 
orthogonal polynomial sequence with the support between zero and infinity for our setup. The number of
polynomials $N$ should be as large as possible, such that Eq.~(\ref{r0}) best describes the subtracted 
continuum function, but cannot be too large in order to avoid the ill-posed problem.

The first term on the left-hand side of Eq.~(\ref{r21}) can be expanded into a power series in $1/x$ trivially.
Because $\Delta\rho_{2n-2}(y)$ decreases quickly enough with the variable $y$, as designed in Eq.~(\ref{sub}), 
the major contribution to its integral arises from a finite range of $y$. It is then justified to expand the 
integral into a series in $1/x$ up to the power $N$ for a sufficiently large $|x|$ by inserting
\begin{eqnarray}
\frac{1}{x-y}=\sum_{i=1}^N \frac{y^{i-1}}{x^i}.
\label{ep0}
\end{eqnarray}
Note that $|x|$ being large enough is only a formal requirement, and does not have a substantial influence 
on the calculation. The right-hand side of Eq.~(\ref{r21}) can be expanded into a power series in $1/x$ 
obviously: both the infrared regulator $P_n$ and the condensate contribution $I_n^{\rm cond}$ appear as power 
series in $1/x$; the exponentials $e^{-y}$ and $e^{-y^4}$, decreasing fast enough with $y$, justify the 
insertion of Eq.~(\ref{ep0}) into the integral. It is then clear that the reformulation of the
power-suppressed logarithm in Eq.~(\ref{i3}) facilitates the expansions of both sides of the dispersion
relation into power series in $1/x$.

Substituting Eqs.~(\ref{r0}) and (\ref{ep0}) into Eq.~(\ref{r21}), and equating the coefficients 
of $1/x^i$ in the power series on the two sides of Eq.~(\ref{r21}), we
arrive at the matrix equation $UA^{(n)}=B^{(n)}$ for each $n$ with the matrix elements
\begin{eqnarray}
U_{i1}&=& r_m^{i-1},\nonumber\\
U_{ij}&=&\int_{0}^\infty dy y^{i-1+\alpha}e^{-y}L_{j-2}^{(\alpha)}(y),\;\;j\ge 2,\label{mm2}
\end{eqnarray}
where $i$ and $j$ run from 1 to $N$. In fact, we have $U_{ij}=0$ for $j>i+1$ owing to the orthogonality
condition in Eq.~(\ref{or0}). The vector 
\begin{eqnarray}
A^{(n)}=(r_f\langle\xi^{2n-2}\rangle\langle\xi^0\rangle,a_{1n}, a_{2n},\cdots,a_{(N-1)n}),\label{an1}
\end{eqnarray}
collects the unknowns associated with the moment $\langle\xi^{2n-2}\rangle$. We point out that the last 
unknown $a_{Nn}$, which also contributes to the coefficient of the term $1/x^N$, has been neglected. In other words, 
$a_{Nn}$ is traded for $\langle\xi^{2n-2}\rangle$, such that the number of the equations is equal to
the number of the unknowns, and the matrix equation is solvable. This approximation is solid, because $a_{jn}$ 
decreases quickly with $j$, as a stable solution for $A^{(n)}$ is attained. That is, a solution, once 
becoming stable, is insensitive to $N$, and whether to keep the last coefficient $a_{Nn}$ is not crucial.

The power expansion on the OPE side gives the coefficient $b_i^{(n)}$ of the term $1/x^i$, 
$n=0,2,4,\cdots$, which are written explicitly as 
\begin{eqnarray}
b^{(n)}_1&=&\lim_{\epsilon\to 0}\frac{1}{\pi}\int_{\epsilon}^\infty dy
\left[{\rm Im}I_n^{\rm pert(1)}e^{-y}+{\rm Im} I_n^{\rm pert(2)}(y\Lambda)e^{-y^4}\right]\nonumber\\
& &-\lim_{\epsilon\to 0}\frac{1}{486\pi^2\epsilon^2}
\left[2(51n+25)-2n\theta(n-2)\right]
\sum_{u,d,s}\frac{\langle g_s^2\bar\psi\psi\rangle^2}{\Lambda^3},\label{opei1}\\
b^{(n)}_2&=&\lim_{\epsilon\to 0}\frac{1}{\pi}\int_{\epsilon}^\infty dyy
\left[{\rm Im}I_n^{\rm pert(1)}e^{-y}+{\rm Im} I_n^{\rm pert(2)}(y\Lambda)e^{-y^4}\right]\nonumber\\
& &-\lim_{\epsilon\to 0}\frac{1}{243\pi^2\epsilon}
\left[2(51n+25)-2n\theta(n-2)\right]
\sum_{u,d,s}\frac{\langle g_s^2\bar\psi\psi\rangle^2}{\Lambda^3}\nonumber\\
& &-\frac{m_u\langle\bar uu\rangle+m_d\langle\bar dd\rangle}{\Lambda^2}
-\frac{1}{12\pi}\frac{\langle \alpha_sG^2\rangle}{\Lambda^2},\label{opei2}\\
b^{(n)}_3&=&\lim_{\epsilon\to 0}\frac{1}{\pi}\int_{\epsilon}^\infty dyy^2
\left[{\rm Im}I_n^{\rm pert(1)}e^{-y}+{\rm Im} I_n^{\rm pert(2)}(y\Lambda)e^{-y^4}\right]\nonumber\\
& &+\lim_{\epsilon\to 0}\frac{1}{243\pi^2}\ln\frac{\epsilon\Lambda}{\mu^2}
\left[2(51n+25)-2n\theta(n-2)\right]
\sum_{u,d,s}\frac{\langle g_s^2\bar\psi\psi\rangle^2}{\Lambda^3}\nonumber\\
& &-\frac{8n+1}{9}\frac{m_u\langle g_s\bar u\sigma TGu\rangle+m_d\langle g_s\bar d\sigma TGd\rangle}{\Lambda^3}
-\frac{n\theta(n-2)}{24\pi^2}\frac{\langle g_s^3fG^3\rangle}{\Lambda^3}\nonumber\\
& &+\frac{4(2n+1)}{81}\frac{\langle g_s\bar uu\rangle^2+\langle g_s\bar dd\rangle^2}{\Lambda^3}
+\frac{C_n}{243\pi^2}\frac{\sum_{u,d,s}\langle g_s^2\bar\psi\psi\rangle^2}{\Lambda^3},\label{opei3}\\
b^{(n)}_i&=&\frac{1}{\pi}\int_{0}^\infty dyy^{i-1}
\left[{\rm Im}I_n^{\rm pert(1)}e^{-y}+{\rm Im} I_n^{\rm pert(2)}(y\Lambda)e^{-y^4}\right],\;\;\;\;i\ge 4.
\label{opei}
\end{eqnarray}
The coefficients $b^{(n)}_{1,2,3}$ of the terms $1/x^{1,2,3}$, respectively, receive the additional 
condensate contributions. The integrals for $b^{(n)}_{i}$, $i\ge 4$, are infrared finite, so the
lower bounds of $y$ have been set to zero. We then define the vector 
\begin{eqnarray}
B^{(n)}=(b^{(2n-2)}_{1},b^{(2n-2)}_{2},\cdots,b^{(2n-2)}_{N}),\;\;\;\;n=1,2,3,\cdots,
\end{eqnarray}
to gather the known inputs from the perturbative and condensate (higher-power) contributions to the OPE.

One can then solve for the vector $A^{(n)}$ straightforwardly via $A^{(n)}=U^{-1}B^{(n)}$ by getting the 
inverse matrix $U^{-1}$. Though an inverse problem is usually ill-posed, i.e., some elements of  $U^{-1}$ may diverge 
when its dimension is sufficiently large, the convergence of Eq.~(\ref{r0}) can be achieved at a finite $N$. 
This convergence, together with the insensitivity of solutions to the variation of $\Lambda$, validate the above 
inverse matrix method. There is no free parameter like the continuum threshold $s_\pi$, and no need to impose 
the discretionary balance between the perturbative and condensate contributions, to apply the Borel 
transformation, and to search for the Borel window in our framework. The full dependence of the dispersion 
relation on $q^2$ up to the power $1/(q^2)^N$ has been utilized to construct the matrix equation. QCD dynamics 
enters through the OPE, which can be derived systematically and rigorously without ambiguity. These merits 
render possible calculating all the moments of the pion LCDA at a scale $\mu$, given the inputs of 
condensates up to certain dimensions.

We stress that the explicit $\mu$ dependence of the dispersion relation for $\langle\xi^0\rangle$,
which defines the normalization of the pion LCDA and is supposed to be always equal to unity, indicates
the incompleteness of the OPE. As postulated in \cite{Zhong:2021epq}, higher-order and 
higher-power corrections to the OPE ought to be included to restore the constant normalization. 
The condition $\langle\xi^0\rangle=1$ was fixed by tuning the continuum threshold
$s_\pi$ in \cite{Zhong:2021epq}, which, however, does not exist in our formalism. We
take an alternative viewpoint in the literature, interpreting the dispersion relation for 
$\langle\xi^0\rangle$ as that for the pion decay constant $f_\pi$ on the premise $\langle\xi^0\rangle=1$.
This viewpoint is equivalent to shifting the $\mu$ dependence of $\langle\xi^0\rangle$ into $f_\pi$.
We thus compute the moments as the ratios
\begin{eqnarray}
\langle\xi^{2n-2}\rangle\equiv\frac{A^{(n)}_{1}}{A^{(1)}_{1}},
\;\;\;\;n=1,2,3,\cdots,
\label{so}
\end{eqnarray}
in which the $\mu$-dependent pion decay constant cancels between the numerator and the denominator. 
We then have the correct normalization $\langle\xi^0\rangle=1$, and resolve the issue about its $\mu$ dependence 
simultaneously.

\subsection{Inverse Matrix Method for LCDA}

One may expect that the $x$ dependence of the pion LCDA can be fully reconstructed, once the information 
of all its moments is available. We warn that it is not the case. Consider the expansion of the pion LCDA 
into a series of Gegenbauer polynomials,
\begin{eqnarray}
\phi_\pi(x)=6x(1-x)\sum_{n=1,2,\cdots}a_{2n-2}^\pi C_{2n-2}^{(3/2)}(2x-1),\label{dom}
\end{eqnarray}
where the coefficients $a_{2n-2}^\pi$ are related to the moments $\langle\xi^{2n-2}\rangle$ up to $n=6$ through
\begin{eqnarray}
a_0^\pi&=&\langle\xi^0\rangle,\nonumber\\
a_2^\pi&=&\frac{7}{12}\left(5\langle\xi^2\rangle-\langle\xi^0\rangle\right),\nonumber\\
a_4^\pi&=&\frac{11}{24}\left(21\langle\xi^4\rangle-14\langle\xi^2\rangle+\langle\xi^0\rangle\right),\nonumber\\
a_6^\pi&=&\frac{5}{64}\left(429\langle\xi^6\rangle-495\langle\xi^4\rangle+135\langle\xi^2\rangle
-5\langle\xi^0\rangle\right),\nonumber\\
a_8^\pi&=&\frac{19}{384} \left(2431 \langle\xi^8\rangle- 4004 \langle\xi^6\rangle
+ 2002 \langle\xi^4\rangle-308 \langle\xi^2\rangle  
 +7\langle\xi^0\rangle\right),\nonumber\\
a_{10}^\pi&=&\frac{23}{1536}\left(29393 \langle\xi^{10}\rangle- 62985 \langle\xi^8\rangle
+ 46410 \langle\xi^6\rangle - 13650 \langle\xi^4\rangle+1365 \langle\xi^2\rangle 
-21\langle\xi^0\rangle\right).\label{cv}
\end{eqnarray}
It is apparent that the coefficients on the right-hand sides of the above relations grow
rapidly with $n$, a feature originating from an ill-posed problem. Then tiny deviations from the true
values of $\langle\xi^{2n-2}\rangle$, due to either theoretical or round-off errors, would destroy the delicate 
cancellation among various terms with huge coefficients in Eq.~(\ref{cv}). 

\begin{figure}
\begin{center}
\includegraphics[scale=0.5]{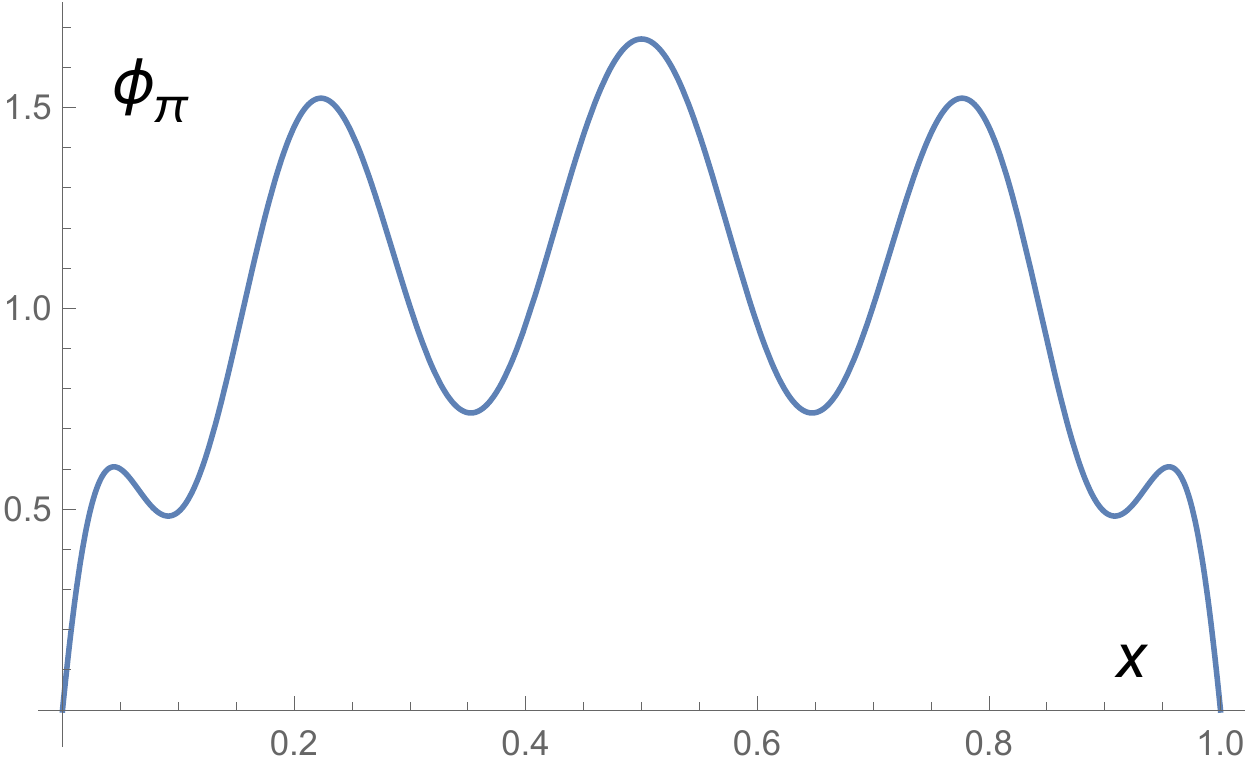}\hspace{0.5cm}

\caption{\label{fig1}
Pion LCDA with the Gegenbauer coefficients in Eq.~(\ref{g00}).}
\end{center}
\end{figure}

Take the results in \cite{Zhong:2021epq} as an example, in which the moments up to $\langle\xi^{10}\rangle$, 
higher than those calculated in the literature, were presented:
\begin{eqnarray}
(\langle\xi^0\rangle,\langle\xi^2\rangle,\langle\xi^4\rangle, \langle\xi^6\rangle,\langle\xi^8\rangle,
\langle\xi^{10}\rangle)|_{\mu=2\,{\rm GeV}}=(1,0.254,0.125,0.077,0.054,0.041).\label{mz}
\end{eqnarray}
They exhibit a satisfactory convergence with $n$, but the corresponding Gegenbauer coefficients
from Eq.~(\ref{cv})
\begin{eqnarray}
(a_0^\pi,a_2^\pi,a_4^\pi, a_6^\pi,a_8^\pi,a_{10}^\pi)|_{\mu=2\,{\rm GeV}}= (1,0.157,0.032,0.035,0.098,-0.046),\label{g00}
\end{eqnarray}
do not. As a consequence, the pion LCDA in Eq.~(\ref{dom}) with the
above Gegenbauer coefficients oscillates strongly with three prominent peaks as displayed in 
Fig.~\ref{fig1}, which seem not to be a reasonable shape. The authors 
of \cite{Zhong:2021epq} then proposed models for the pion LCDA, and fitted the involved parameters to the 
moments in Eq.~(\ref{mz}), instead of getting the Gegenbauer coefficients in Eq.~(\ref{g00}). The above 
discussion exemplifies that the conversion from the moments to the Gegenbauer coefficients goes out of 
control quickly with $n$, and that acquiring the $x$ dependence of a LCDA is a challenging subject. 
Nevertheless, we will evaluate the moments of the pion LCDA in the next section to 
demonstrate the potential of our method for accessing the full $x$ dependence of a LCDA. 

To overcome the aforementioned difficulty, we extend the formalism established in the previous subsection.
The idea is to avoid the delicate cancellation in Eq.~(\ref{cv}) by searching for stable solutions 
of the Gegenbauer coefficients directly. We define a matrix $V$ with the elements
\begin{eqnarray}
V_{kn}&=&6\int_0^1 dx x(1-x)(2x-1)^{2n-2}C_{2k-2}^{(3/2)}(2x-1),\label{vm}
\end{eqnarray}
where $V_{kn}=0$ for $k>n$ due to the orthogonality of the Gegenbauer polynomials. It is trivial to check that
the elements of the inverse matrix $V^{-1}$ contain the coefficients on the right-hand side of Eq.~(\ref{cv}). 
Namely, the matrix $V$ is responsible for the conversion between the moments and the Gegenbauer coefficients.  
Next we rewrite the matrix $a$ formed by the coefficients $a_{jn}$ in the expansion of the subtracted continuum 
function in Eq.~(\ref{r0}) as $a=\tilde a V$, so that all the unknowns are grouped into a single matrix 
\begin{eqnarray}
A&=&\left( \begin{array}{cccc}
    r_f\langle\xi^0\rangle a_0^\pi&r_f\langle\xi^0\rangle a_2^\pi & \cdots &r_f\langle\xi^0\rangle a_{2N-2}^\pi\\
    \tilde a_{11} & \tilde a_{12} &\cdots &\tilde a_{1N}\\
    \cdots &\cdots &\cdots &\cdots \\
    \tilde a_{(N-1)1} & \tilde a_{(N-1)2} &\cdots &\tilde a_{(N-1)N}
   \end{array} \right).\label{aun}
\end{eqnarray}
It is then straightforward to construct the matrix equation $UAV=B$ by repeating the procedure in the 
previous subsection, where the elements of the matrix $U$ are the same as in Eq.~(\ref{mm2}), the unknown 
matrix $A$ contains the Gegenbauer coefficients $a_{2n-2}^\pi$, and the matrix 
$B$ collects the inputs with the elements $B_{in}=B_i^{(n)}=b_i^{(2n-2)}$ in Eqs.~(\ref{opei1})-(\ref{opei}). 
Similarly, the last row of $\tilde a$, i.e., $\tilde a_{N1}$, $\tilde a_{N2}$, 
$\cdots$, $\tilde a_{NN}$ have been neglected and traded for the first row of $A$ in order to have
equal numbers of equations and unknowns.  We will justify this approximation by showing that 
the last row of $\tilde a$ are indeed negligibly small in the next section. According to the argument
for Eq.~(\ref{so}), we consider the ratios in this work
\begin{eqnarray}
a_{2n-2}^\pi\equiv\frac{A_{1n}}{A_{11}},\;\;\;\;n=1,2,3,\cdots,
\label{soa}
\end{eqnarray}
which follows the interpretation with the constant normalization $\langle\xi^0\rangle=a_0^\pi=1$.

Naively, one can get a solution of $A=U^{-1}BV^{-1}$ through the inverse matrices $U^{-1}$ and $V^{-1}$, 
where the effect of $V^{-1}$ is simply to organize the OPE inputs $B$ for the moments into
$BV^{-1}$ for the Gegenbauer coefficients. However, the above inverse matrix method may not lead to
stable solutions for the Gegenbauer coefficients due to the ill-posed nature, especially when inputs are 
not accurate enough. A standard and popular resolution to this issue is to apply the Tikhonov regularization,
i.e., to add an additional constraint which smears the fluctuation caused by imperfect cancellation among 
various terms in Eq.~(\ref{cv}). The inverse matrix $V^{-1}$ diverges more seriously than $U^{-1}$ with $N$:
the maximal elements of $V^{-1}$ and $U^{-1}$ for the dimension $N=18$ are 
of $O(10^{10})$ and $O(10^2)$, respectively. This difference explains why stable solutions for the moments, 
which require only $U^{-1}$, can be found at a finite $N$, and why an explicit regularization is necessary
for suppressing the divergence in $V^{-1}$. Therefore, we propose the modified matrix equation
\begin{eqnarray}
UA(V+\lambda H)=B,
\end{eqnarray}
where $\lambda$ ($H$) denotes a regularization parameter (matrix). There is freedom
to choose $H$, but not any $H$ works for stabilizing solutions. 
We will show that a simple choice $H=I$, $I$ being the unity matrix, serves the purpose well,
and look for stable solutions of $A=U^{-1}B(V+\lambda I)^{-1}$, which are insensitive to the 
regularization parameter $\lambda$.

The scale $\mu$ is set to the Borel mass $M$ in \cite{Zhong:2021epq}, and the moments for $\mu=1$ GeV and 
2 GeV were then extracted through the QCD evolution. The value of $\mu$ in the OPE input in Eq.~(\ref{opei3}) 
can be chosen arbitrarily. We will determine the pion LCDAs at $\mu=2$ GeV and $\mu=1.5$ GeV, and examine 
whether these two results are consistent with the known QCD evolution that connects the Gegenbauer coefficients 
at different scales. It will be observed that the LCDA solved from the dispersion relation with the OPE
inputs at $\mu=1.5$ GeV agrees with the one derived by evolving the Gegenbauer coefficients at $\mu=2$
GeV to $\mu=1.5$ GeV. This agreement hints that our formalism is compatible with the QCD evolution.

\section{NUMERICAL ANALYSIS}

\subsection{Testing the Formalism with Mock Data}

We first demonstrate that correct solutions can be obtained in our method, given the mock data 
generated from a sample LCDA in Eq.~(\ref{dom}) with the Gegenbauer coefficients
\begin{eqnarray}
(a_0^\pi,a_2^\pi, a_4^\pi,a_6^\pi, a_8^\pi,a_{10}^\pi,\cdots) =(1,0.20,-0.15,0.10,0,0,\cdots),\label{ts2}
\end{eqnarray}
and from a set of sample continuum functions
\begin{eqnarray}
\Delta\rho_{2n-2}(y)&=&y e^{-ny},\;\;\;\;n=1,2,\cdots.\label{ts1}
\end{eqnarray}
The moments $\langle\xi^{2n-2}\rangle$ are computed according to Eq.~(\ref{mel}) and then,
together with Eq.~(\ref{ts1}), substituted into the left-hand side of Eq.~(\ref{r21}), whose power expansion 
in $1/x^i$ produces the elements $B_i^{(n)}$ of the input matrix $B^{(n)}$,
\begin{eqnarray}
B_i^{(n)}=r_m^{i-1}\int_{0}^1 dy (2y-1)^{2n-2}\phi_\pi(y)+
\int_{0}^\infty dy y^ie^{-ny}.\label{bn1}
\end{eqnarray}
Here the factor $r_f$ has been omitted for simplicity, which cancels in Eq.~(\ref{so}), the pion mass takes 
the value $m_\pi = 139.57$ MeV \cite{PDG}, and the transition scale is set to $\Lambda=1$ GeV$^2$. Another 
choice $\Lambda=2$ GeV$^2$ causes no change to the results. We evaluate the $N\times N$ matrix 
$U$ following Eq.~(\ref{mm2}), where $N$ is related to the maximal degree of the generalized Laguerre 
polynomial in the expansion in Eq.~(\ref{r0}). In principle, a true solution can be approached to by 
increasing the number of polynomials $N$, since the difference 
between the true and approximate solutions is suppressed by a power $1/x^{N+1}$. However, $N$ cannot be
too large in a practical application, otherwise the approximate solution would deviate from the true one 
due to the generic nature of an ill-posed inverse problem as mentioned before.

For each $n$, we increase the dimension of the matrix $U$, and derive the solutions $A^{(n)}=U^{-1}B^{(n)}$
according to Sec.~IIB. It is found that the solutions become stable gradually as $N$ enlarges, 
vary only by about 0.1\% within the interval $N=[17,21]$, and begin to change significantly as $N\ge 22$, 
implying that the inverse matrix $U^{-1}$ is out of control.
The  solutions of $A^{(n)}$ read up to $n=7$ for $N=19$
\begin{eqnarray}
& &(\langle\xi^0\rangle,a_{11}, a_{21},a_{31},\cdots,a_{(17)1},a_{(18)1})=
(1, 1, \sim 0,\sim 0, \cdots, \sim 0, \sim 0),\nonumber\\
& &(\langle\xi^2\rangle,a_{12}, a_{22},a_{32},\cdots,a_{(17)2},a_{(18)2})=
(0.2686, 0.2500, 0.1250, 0.0625,\cdots, -5.0\times 10^{-6}, -3.6\times 10^{-6}),\nonumber\\
& &(\langle\xi^4\rangle,a_{13}, a_{23},a_{33},\cdots,a_{(17)3},a_{(18)3})=
(0.1159, 0.1110, 0.0740, 0.0493, \cdots, 9.0\times 10^{-5}, 3.6\times 10^{-5}),\nonumber\\
& &(\langle\xi^6\rangle,a_{14}, a_{24},a_{34},\cdots,a_{(17)4},a_{(18)4})=
(0.0642, 0.0621, 0.0465, 0.0347, \cdots, 2.7\times 10^{-4}, 1.1\times 10^{-4}),\nonumber\\
& &\langle\xi^8\rangle=0.0417,\;\;\;\;\langle\xi^{10}\rangle=0.0300,\;\;\;\;
\langle\xi^{12}\rangle=0.0232,\label{t1}
\end{eqnarray}
where the notation $\sim 0$ represents a value with magnitude being smaller than $10^{-10}$,
and the solutions for the coefficients $a_{jn}$ in Eq.~(\ref{r0}) for $n=5,6,7$ are not shown explicitly. 
The monotonically decreasing sequences $a_{jn}$ in $j$ and the smallness of the last elements 
$a_{(18)n}$ support the neglect of $a_{(19)n}$ in the construction of the matrix equation.
Compared to the true solutions, 
\begin{eqnarray}
(\langle\xi^0\rangle,\langle\xi^2\rangle,\langle\xi^4\rangle,\langle\xi^6\rangle,
\langle\xi^8\rangle,\langle\xi^{10}\rangle,\langle\xi^{12}\rangle)=
(1,0.2686,0.1158,0.0638,0.0408,0.0288,0.0217),\label{mm3}
\end{eqnarray}
from the sample LCDA with the Gegenbauer coefficients in Eq.~(\ref{ts2}),
the moments, especially the first few, have been reproduced perfectly.

\begin{figure}
\begin{center}
\includegraphics[scale=0.4]{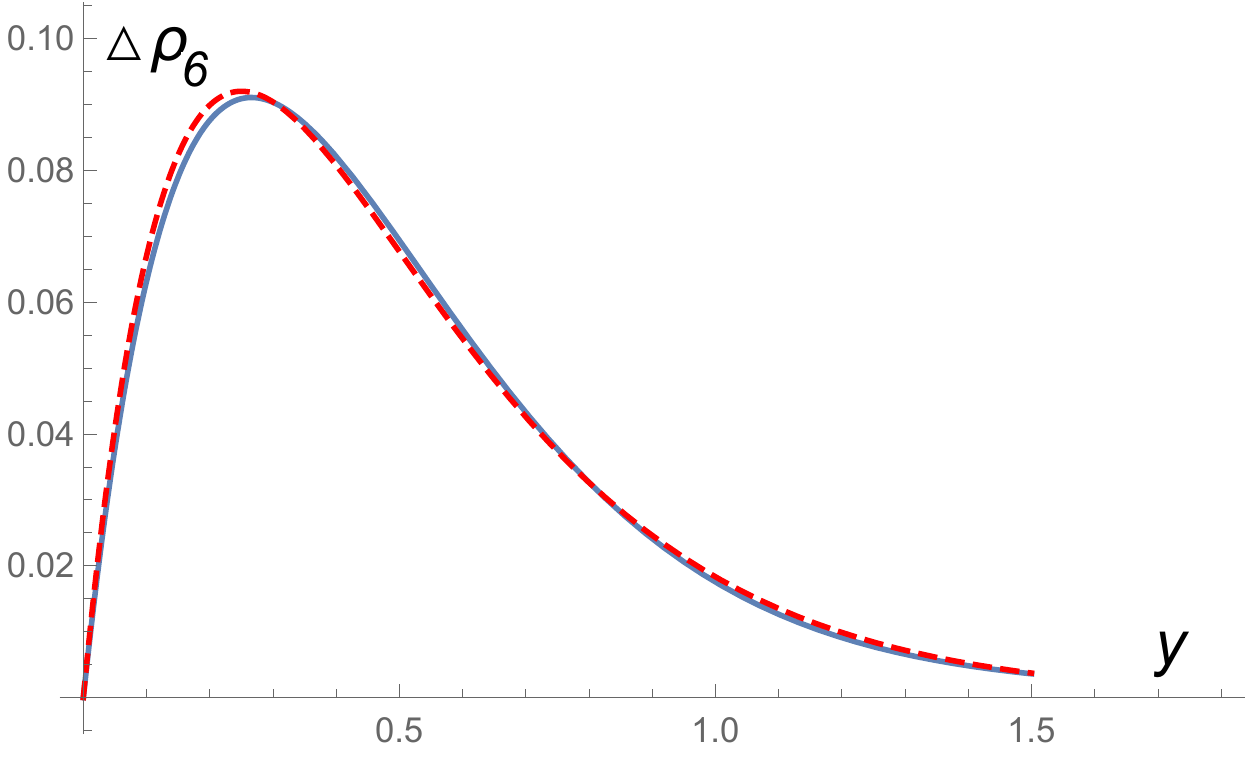}\hspace{0.5cm}
\includegraphics[scale=0.4]{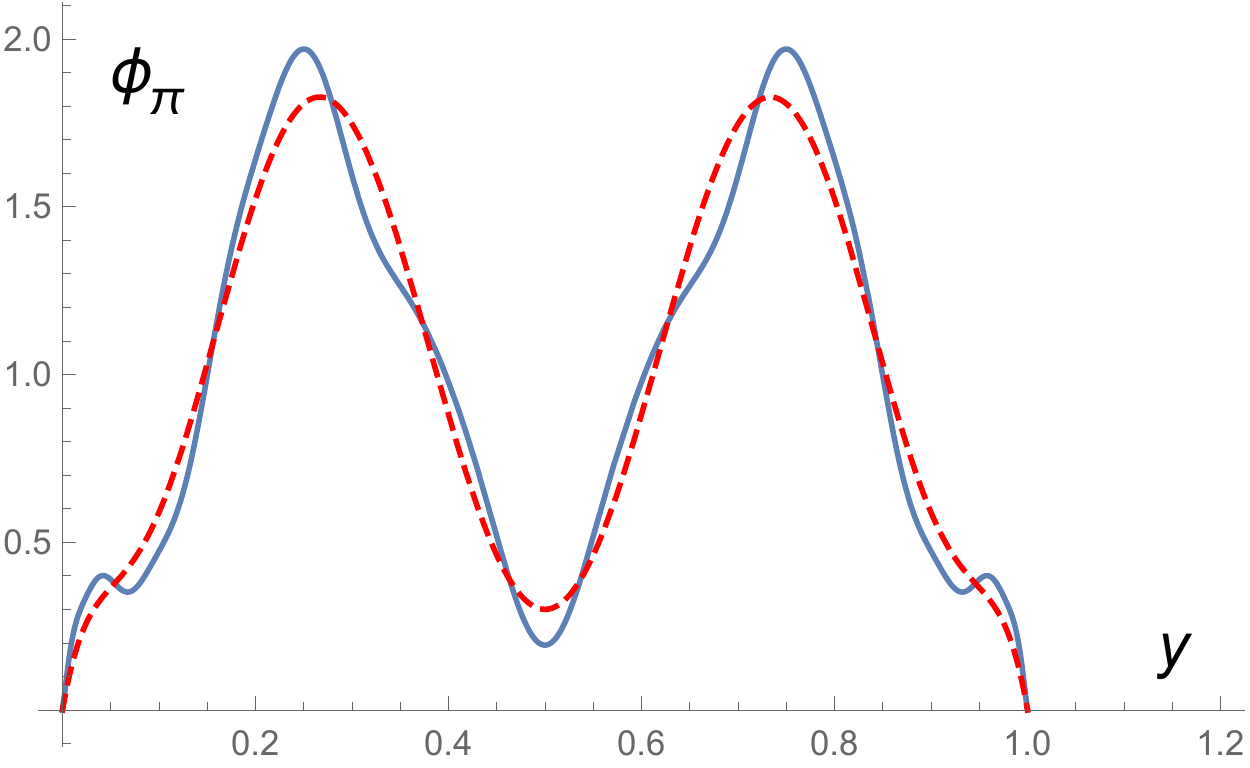}\hspace{0.5cm}
\includegraphics[scale=0.4]{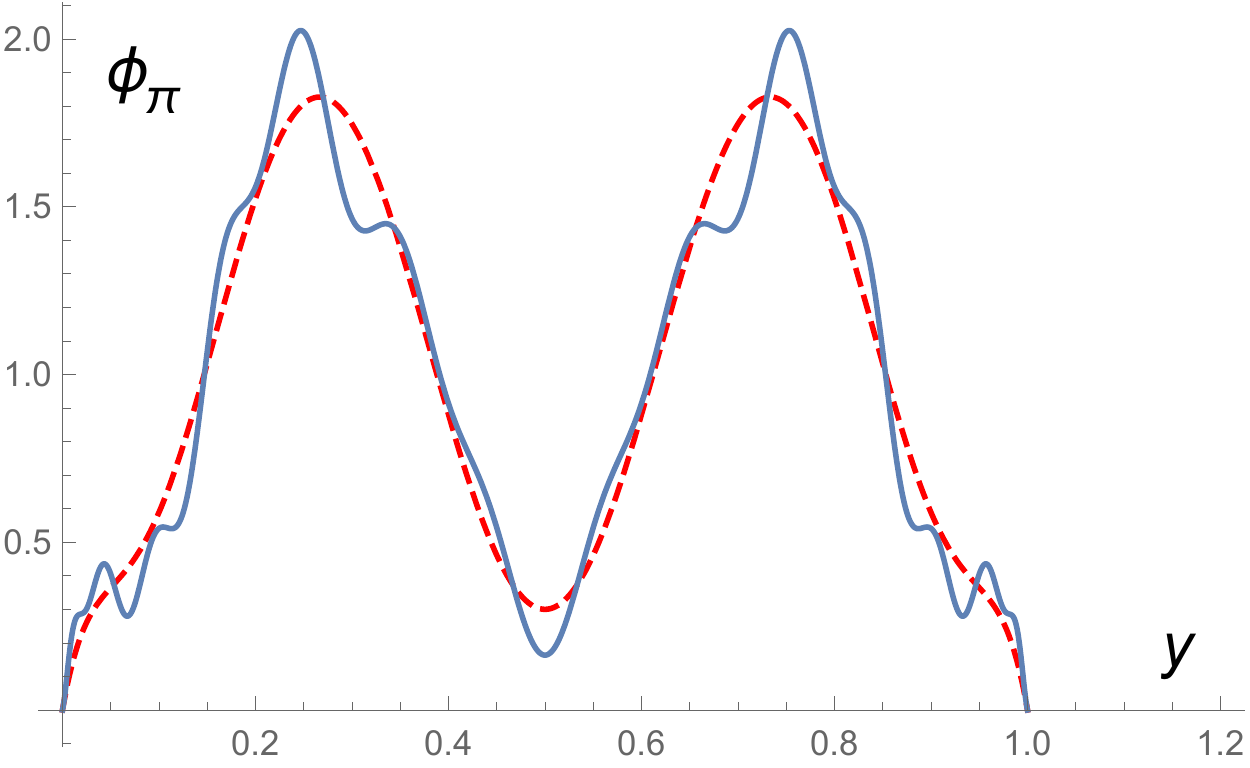}

(a) \hspace{5.0 cm} (b)\hspace{5.0 cm} (c)
\caption{\label{fig2}
(a) Solution of $\Delta\rho_{6}(y)$ (solid line) and the input one  $y e^{-4y}$ (dashed line).  
(b) Solution of $\phi_\pi(y)$ with $N=16$ (solid line) and the input one (dashed line).
(c) Solution of $\phi_\pi(y)$ with $N=17$ (solid line) and the input one (dashed line).}
\end{center}
\end{figure}

The solution for the continuum function $\Delta\rho_{6}(y)$ and the input one $y e^{-4y}$ in Eq.~(\ref{ts1}) 
are compared in Fig.~\ref{fig2}(a), whose consistency confirms the good quality of the solutions.
If the solution for $\Delta\rho_{2}(y)$ and the input one $y e^{-2y}$ are plotted, 
their curves overlap exactly. Even so, the above seemly accurate results cannot reproduce the $x$ dependence 
of the sample LCDA: the moments in Eq.~(\ref{t1}) lead to the Gegenbauer
coefficients $a_{2n-2}^\pi$ through the relations in Eq.~(\ref{cv}),
\begin{eqnarray}
(a_0^\pi,a_2^\pi,a_4^\pi,a_6^\pi, a_8^\pi, a_{10}^\pi,a_{12}^\pi)=
(1,0.2001, -0.1496, 0.1119, 0.0306, -0.0233, 0.2339),\label{ga1}
\end{eqnarray} 
which deviate from the true values in Eq.~(\ref{ts2}) more at higher $n$. In particular, the magnitude of $a_{12}^\pi$ 
becomes larger than $a_2^\pi$, reflecting the ill-posed nature of the subject. One may enhance the 
precision of the calculation by keeping more digits of the numbers, but the appearance of the divergent 
Gegenbauer coefficients is just deferred to even higher $n$, and the problem is not completely resolved. This is 
the same difficulty as elucidated in terms of the moments in \cite{Zhong:2021epq} in Sec.~IIC.

We turn to the method developed in Sec.~IIC, and retrieve the Gegenbauer coefficients from the mock data.
The sample LCDA and continuum functions in Eq.~(\ref{ts1}) generate the input matrix $B$ with the elements 
$B_{in}=B_i^{(n)}$ the same as in Eq.~(\ref{bn1}). We get the matrices $U$ and $V$ up to the dimension $N\times N$ 
following Eqs.~(\ref{mm2}) and (\ref{vm}), respectively, and the solutions $A=U^{-1}BV^{-1}$, whose first 
rows give the Gegenbauer coefficients. Because the data are precise, it turns out that no regularization is 
needed to stabilize the solutions. We find that the LCDAs constructed from the obtained Gegenbauer 
coefficients are stable after $N$ reaches 14, and start to oscillate as $N\ge 17$. The resultant 
LCDA $\phi_\pi(y)$ for $N=16$ and the sample LCDA are compared in Fig.~\ref{fig2}(b), which 
match each other roughly. The similarity is justified by the approximate equality between the first few moments 
for the obtained LCDA and Eq.~(\ref{mm3}). The shape of the obtained $\phi_\pi(y)$ is a 
bit irregular for $N=17$ in Fig.~\ref{fig2}(c), which signals the divergent behavior of the inverse matrices 
$U^{-1}$ and $V^{-1}$, and will become strongly oscillatory as $N$ increases further. 
The Gegenbauer coefficients corresponding to the solution for $N=16$
\begin{eqnarray}
& &(a_0^\pi,a_2^\pi,a_4^\pi,a_6^\pi, a_8^\pi, a_{10}^\pi,a_{12}^\pi,\cdots,a_{28}^\pi,a_{30}^\pi)\nonumber\\
&=&(1,0.2000, -0.1472, 0.1212, 0.0335, -0.0059, -0.0098,\cdots,-0.0029, 0.0013),\label{ga3}
\end{eqnarray}
are all under control, a tendency quite distinct from that of Eq.~(\ref{ga1}). The coefficient $a_2^\pi$ is 
precisely reproduced, but $a_6^\pi$ exceeds the true value in Eq.~(\ref{ts2}) by 20\%. The larger discrepancy 
between the Gegenbauer coefficients with smaller magnitude, such as $a_{10}^\pi$ and $a_{12}^\pi$, is not a 
surprise. We confirm that the corresponding sequences of $\tilde a_{jk}$ associated with the sample continuum 
functions also converge well in $j$, and the last row in the unknown matrix $\tilde a$ is negligible. For 
$N=17$, the last two Gegenbauer coefficients take the values $a_{30}^\pi=0.0050$ and $a_{32}^\pi=0.0035$ with 
convergence slightly worse than in Eq.~(\ref{ga3}).

\begin{figure}
\begin{center}
\includegraphics[scale=0.4]{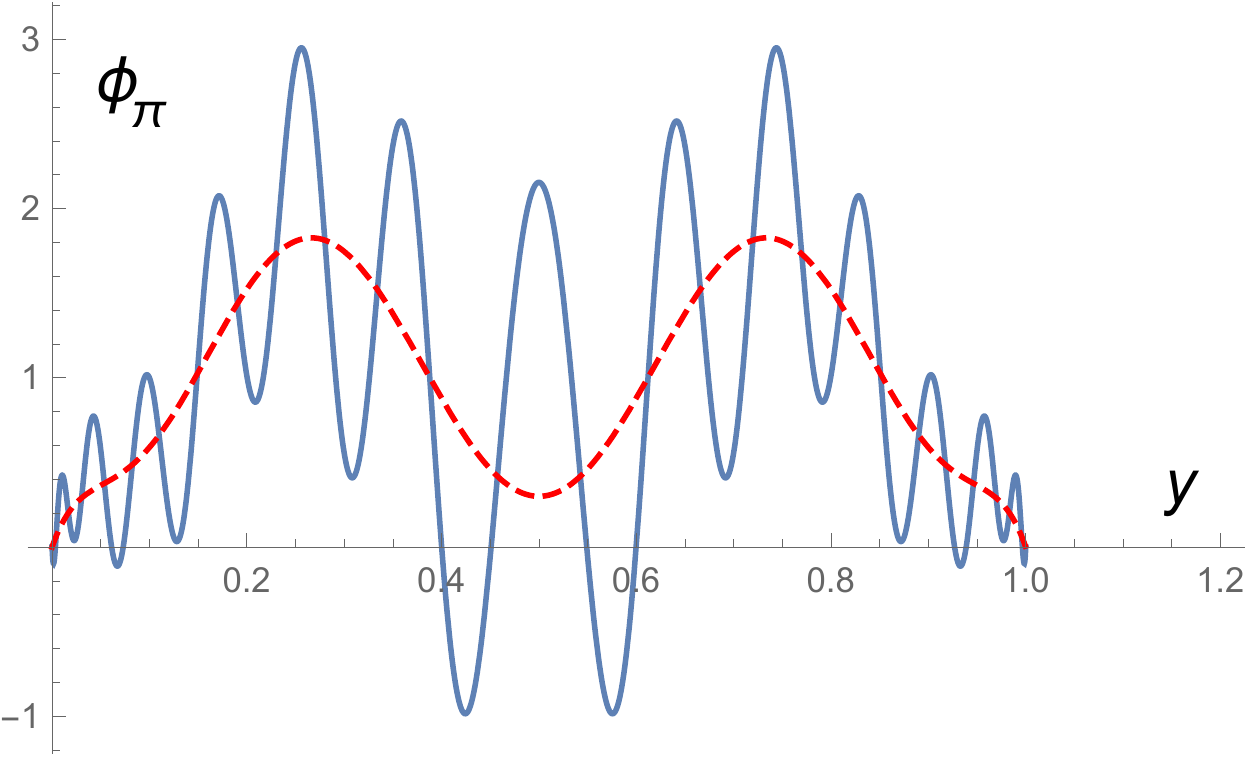}\hspace{0.5cm}
\includegraphics[scale=0.4]{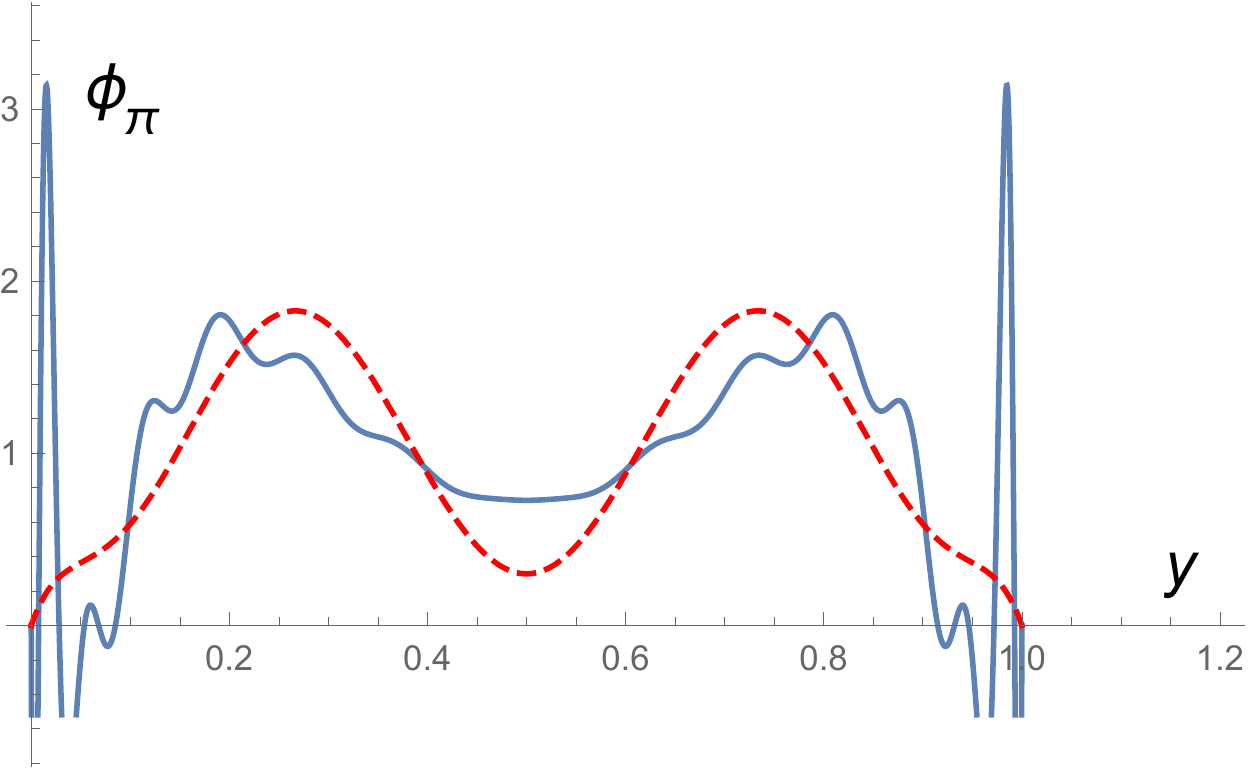}\hspace{0.5cm}
\includegraphics[scale=0.4]{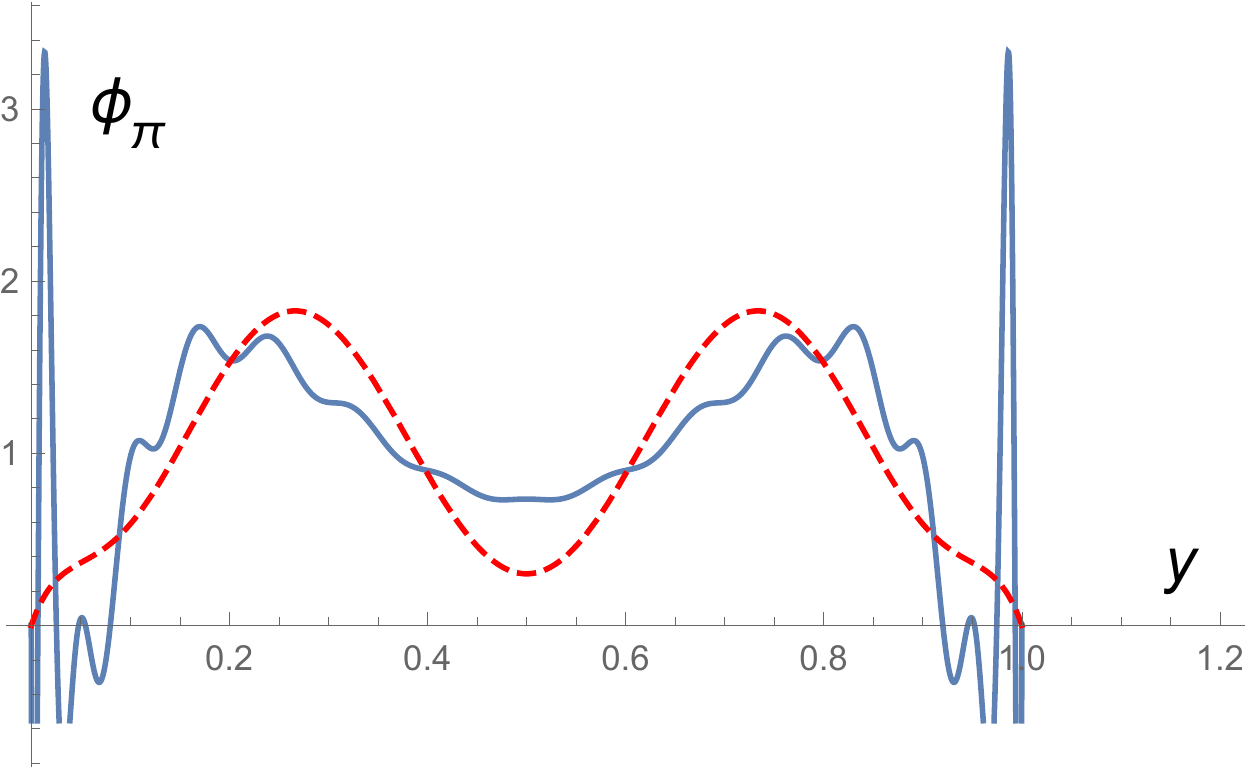}\hspace{0.5cm}

(a) \hspace{5.0 cm} (b)\hspace{5.0 cm} (c)
\caption{\label{fig3}
Solutions of $\phi_\pi(y)$ for (a)  $\lambda=0$ with $N=16$ (solid line) and the input one (dashed line),
(b) $\lambda=0.0054$ with $N=15$ (solid line) and the input one (dashed line), and
(c) $\lambda=0.0058$ with $N=16$ (solid line) and the input one (dashed line).}
\end{center}
\end{figure}

At last, we add a tiny fluctuation to the input matrix $B$, enhancing the element $B_{22}$ by 0.05\%,
which suffices to destroy the stability of the solutions: Figure~\ref{fig3}(a) indicates that the solution
of $\phi_\pi(y)$ for $N=16$ without the regularization exhibits violent oscillations compared to 
the one in Fig.~\ref{fig2}(b), and differs from the sample LCDA completely. We then switch on the regularization, 
and investigate the behavior of the solutions with the parameter $\lambda$. It is observed that the Gegenbauer 
coefficients converge and the resultant $\phi_\pi(y)$ in Fig.~\ref{fig3}(b) is insensitive to the variation of 
$\lambda$ around $\lambda=0.0054$ for $N=15$. The shape of $\phi_\pi(y)$ has been greatly improved 
relative to the one in Fig.~\ref{fig3}(a), and gets closer to that of the sample LCDA. For $N=16$, the stability 
interval in $\lambda$ moves toward somewhat larger values around $\lambda=0.0058$, and the obtained $\phi_\pi(y)$ 
is similar to the one for $N=15$, implying the stability of the solutions with the increase of the dimension $N$ 
of the matrices $U$ and $V$. The Gegenbauer coefficients corresponding to the solution for $N=16$ are given by
$a_2^\pi=0.1980$, $a_4^\pi=-0.1289$, $a_6^\pi=0.0597$,.... It is hard to tell how much their deviation  
from those in Eq.~(\ref{ts2}) is attributed to the added fluctuation, and how much originates from the
inverse matrix method. The lessons from the above analysis of the mock data include that a LCDA can be 
uncovered to some extent in our formalism even with fluctuations in inputs, and that the direct evaluation of 
the Gegenbauer coefficients is more promising than going through the moments for exploring the $x$ dependence
of a LCDA.

\subsection{Moments of the pion LCDA}

After testing our formalism with the mock data, we apply it to the study of the realistic pion 
LCDA, starting with the determination of the moments from the OPE inputs. The parameter $\epsilon$ 
is set to $\epsilon=10^{-5}$, which is small enough. We have checked that the choice $\epsilon=10^{-6}$ 
changes solutions for the moments at the level of $10^{-4}$, much lower than theoretical uncertainties 
from other sources. To compare our results with those in \cite{Zhong:2021epq}, we take the same values 
for the following condensates \cite{SN15,CK00} with the evolution in the scale $\mu$
\cite{YHH,HY94,LWZ},
\begin{eqnarray}
& &m_u \langle\bar{u}u\rangle + m_d \langle\bar{d}d\rangle
= -(1.651 \pm 0.003) \times 10^{-4} \;{\rm GeV}^4,\nonumber\\
& &\langle g_s\bar q q\rangle^2 =(2.082^{+0.734}_{-0.697})\times 10^{-3}
\left[\frac{\alpha_s(\mu)}{\alpha_s(2\;{\rm GeV)}}\right]^{-4/\beta_0}\;{\rm GeV}^6,\nonumber\\
& &\sum_{u,d,s} \langle g_s^2\bar{\psi}\psi\rangle^2 = (2+r_c^2)\langle g_s^2\bar q q\rangle^2,\;\;
\langle g_s^2\bar q q\rangle^2=(7.420^{+2.614}_{-2.483})\times 10^{-3}\; {\rm GeV}^6,\nonumber\\
& &\langle\alpha_s G^2\rangle = 0.038 \pm 0.011,\;{\rm GeV}^4,\;\;\;\;
\nonumber\\
& &m_u \langle g_s\bar{u}\sigma TGu\rangle + m_d \langle g_s\bar{d}\sigma TGd\rangle
= -(1.321 \pm 0.033) \times 10^{-4}
\left[\frac{\alpha_s(\mu)}{\alpha_s(2\;{\rm GeV)}}\right]^{14/(3\beta_0)}\; {\rm GeV}^4,
\label{mqq}
\end{eqnarray}
for $q=u$ or $d$ and $\beta_0 = 11 - 2n_f/3$, $n_f=4$ being the number of active quark flavors. 
For simplicity, we consider the one-loop running coupling constant $\alpha_s(\mu)$ with
the QCD scale $\Lambda_{\rm QCD}=0.22$ GeV.
Those condensates without the evolution factors are scale independent; namely, their values
are the same as at $\mu=2$ GeV. In fact, only 
the evolution of the four-quark condensate $\langle g_s\bar q q\rangle^2$ matters. As summarized 
in \cite{Harnett:2021zug}, the ratio $r_c\equiv \langle\bar{s}s\rangle/ \langle\bar{q}q\rangle$ 
derived in the literature varies in a wide range, from 0.4 to 1.2. 
Here we also follow \cite{Zhong:2021epq}, taking the average $r_c=0.74 \pm 0.03$ presented in \cite{QUACON}.

The triple-gluon condensate $\langle g_s^3fG^3\rangle$ has been estimated in the single-instanton model 
\cite{SVZ,NS80,RRY}, in lattice QCD \cite{PV90}, and via the sum rules for charmonium systems 
\cite{SN10}. The results differ dramatically as having been noticed in \cite{Narison:2018dcr}: the first 
two are opposite in sign, and the last one has a magnitude about several times larger. It has been observed 
that this input impacts the predictions for glueball masses, so its values can be discriminated by the associated 
low energy theorem \cite{Li:2021gsx}. Since glueball states have not yet been identified unambiguously, the 
value of $\langle g_s^3fG^3\rangle$ is still uncertain. The dispersion relation for the moment
$\langle\xi^0\rangle$ does not depend on $\langle g_s^3fG^3\rangle$ as indicated in Eq.~(\ref{con6}), 
so the corresponding prediction is free of this ambiguity. The higher moments depend on $\langle g_s^3fG^3\rangle$,
and we find no stable solution for the moment $\langle\xi^2\rangle$ according to Eq.~(\ref{so}) with the input 
$\langle g_s^3fG^3\rangle = 0.045$ GeV$^6$ in \cite{Zhong:2021epq,CK00}. 
For a stable solution to exist, $\langle g_s^3fG^3\rangle$ must be sizable enough to compensate the negative 
contribution from the four-quark condensates $\langle g_s\bar q q\rangle^2$ and 
$\langle g_s^2\bar q q\rangle^2$ at the same power of $1/(q^2)^3$ in Eq.~(\ref{con6}). Hence, we choose the estimate 
in \cite{SN10}, 
\begin{eqnarray}
\langle g_s^3fG^3\rangle = (8.2\pm 1.0)\; {\rm GeV}^2 \times\langle\alpha_sG^2\rangle,
\label{gg3}
\end{eqnarray}
which leads to a result for $\langle\xi^2\rangle$ close to the one in \cite{Zhong:2021epq}.

\begin{figure}
\begin{center}
\includegraphics[scale=0.4]{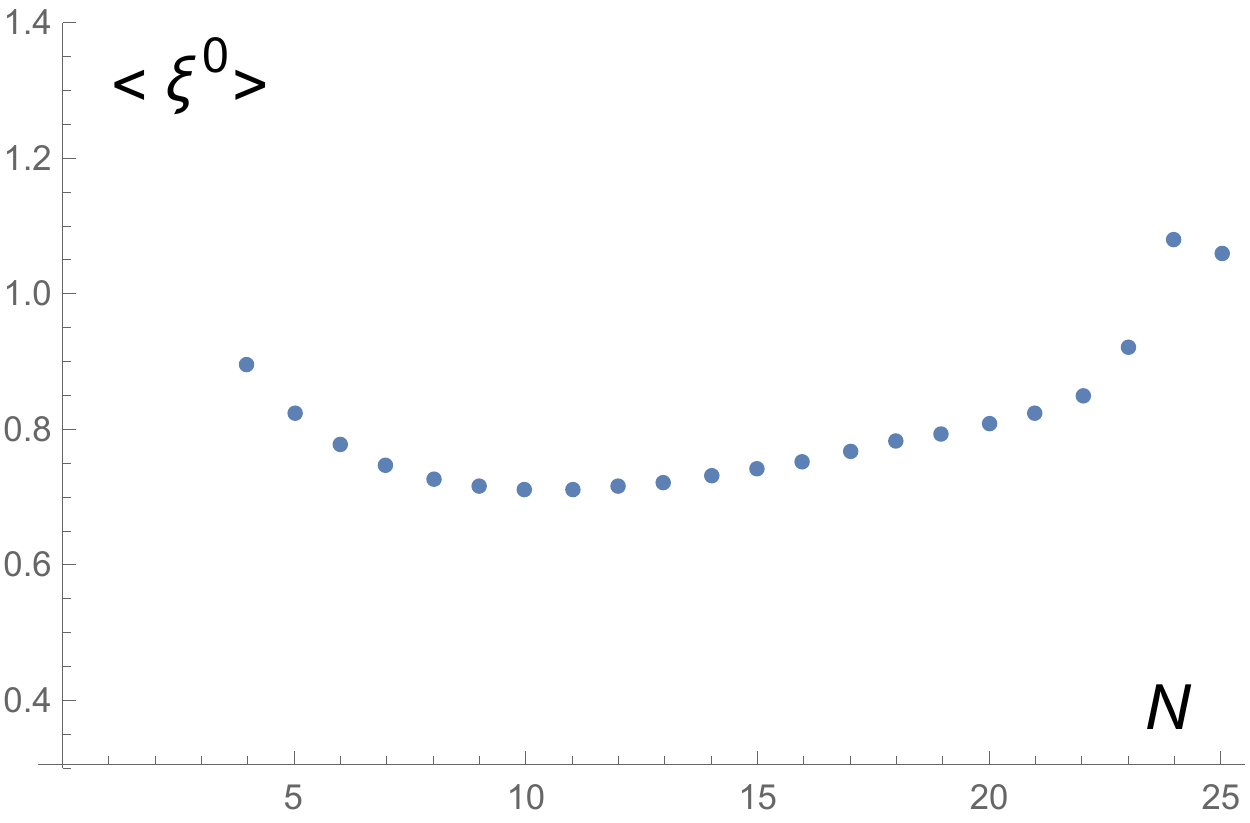}\hspace{0.5cm}
\includegraphics[scale=0.4]{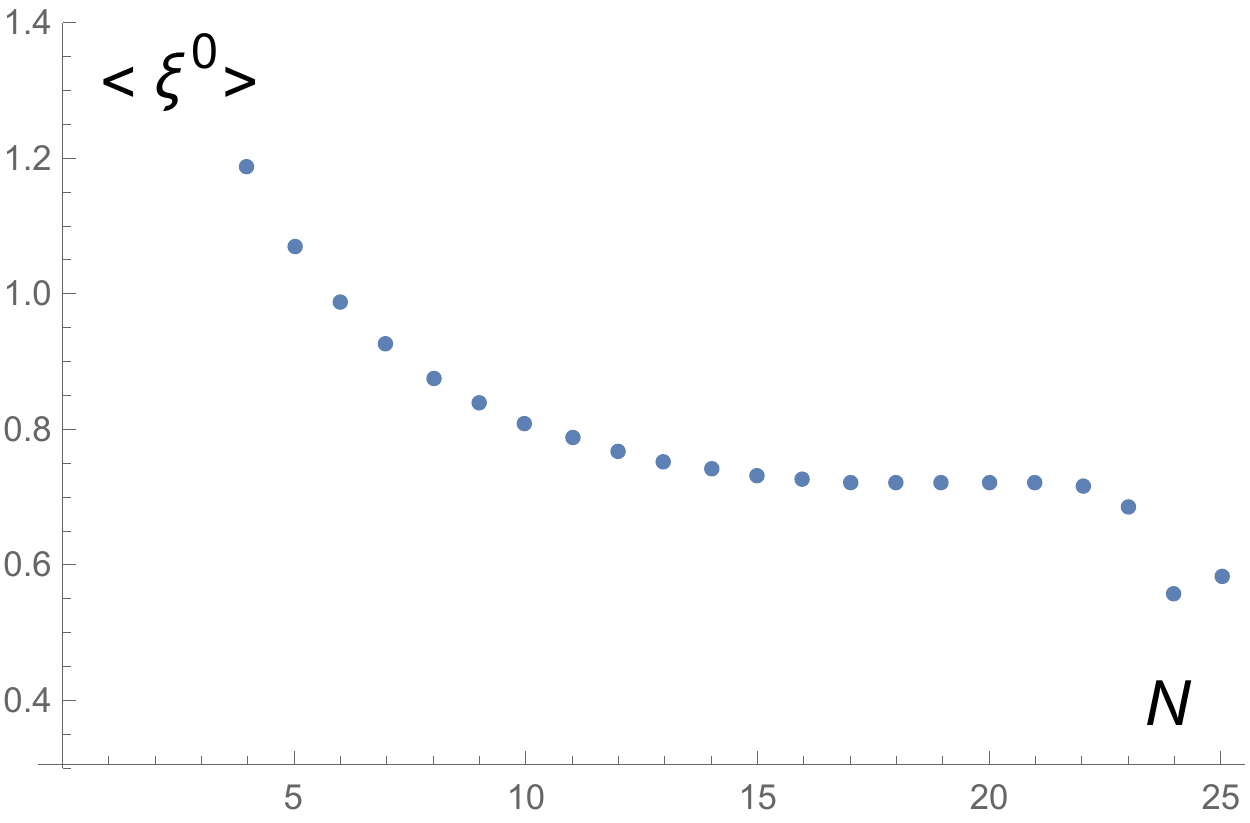}

(a) \hspace{5.0 cm} (b)
\caption{\label{fig4}
Dependencies of $\langle\xi^0\rangle$ on $N$ for (a) $\Lambda=2.0$ GeV$^2$ and (b)  $\Lambda=3.7$ GeV$^2$.}
\end{center}
\end{figure}

We first analyze the zeroth moment $\langle\xi^0\rangle$ from $A^{(1)}=U^{-1}B^{(1)}$ with $B^{(1)}$ in 
Eqs.~(\ref{opei1})-(\ref{opei}) as a demonstration. To take into account the dimension-six condensates, the 
dimension of the matrix $U$ should be greater than three, so we start with $N=4$, and increase $N$ one by 
one to search for a convergent expansion in Eq.~(\ref{r0}) for a given transition scale $\Lambda$. 
When the convergence is attained, the solution for $A^{(1)}$, including its first component, becomes relatively
stable with respect to the variation of $N$. The value of $\langle\xi^0\rangle$ for the given $\Lambda$ is 
then obtained via $\langle\xi^0\rangle=\sqrt{A^{(1)}_{1}/r_f}$,
as the pion decay constant $f_\pi=130.41$ MeV \cite{PDG} is known.
Figures~\ref{fig4}(a) and \ref{fig4}(b) exhibit the dependencies of $\langle\xi^0\rangle$ on $N$ for 
$\Lambda=2$ GeV$^2$ and 3.7 GeV$^2$, respectively. It is found in Fig.~\ref{fig4}(a)
that the zeroth moment decreases from $N=4$, reaches a minimum at $N=10$, and then increases
with $N$ due to the growth of $U^{-1}$ at high $N$. As $N$ exceeds 22, the matrix elements 
of $U^{-1}$ begin to go out of control due to the ill-posed nature, such that the zeroth
moment changes significantly. As $\Lambda$ increases, the range of $N$, in which the zeroth moment stays around 
the minimum, becomes wider, implying better convergence of the expansion in Eq.~(\ref{r0}). The 
flatness of the curve for the zeroth moment, the convergence of the expansion, and the stability of the 
solution at $\Lambda=3.7$ GeV$^2$ are remarkable as $N>16$ in Fig.~\ref{fig4}(b), before the curve 
descends and oscillates at $N=23$. Note that there is a minimum of $\langle\xi^0\rangle$ at $N=19$ in 
Fig.~\ref{fig4}(b), though it is not visible in the almost flat plateau.

To justify the neglect of the coefficient $a_{N1}$ in the construction of the matrix equation,
we show the solutions of $A^{(1)}$ corresponding to the minima located at $N=10$ in Fig.~\ref{fig4}(a)
and located at $N=19$ in Fig.~\ref{fig4}(b),
\begin{eqnarray}
A^{(1)}|_{N=10}&=&(0.0043, 0.0210, 0.0085,\cdots, 1.7\times 10^{-4}, 4.9\times 10^{-5}),\label{A1}\\
A^{(1)}|_{N=19}&=&(0.0024, 0.0229, 0.0103, \cdots, 1.9\times 10^{-5}, 5.0\times 10^{-6}),\label{A2}
\end{eqnarray}
where the first components give the solutions for $\langle\xi^0\rangle$. 
The small ratio $a_{91}/a_{11}=4.9\times 10^{-5}/0.0210\approx 0.002$ and $a_{91}$ being a quarter 
of $a_{81}$ in Eq.~(\ref{A1}) support that the unknown coefficient $a_{(10)1}$ can be neglected
safely. The tiny ratio $a_{(18)1}/a_{11}=5.0\times 10^{-6}/0.0229\approx 2\times 10^{-4}$ and 
$a_{(18)1}$ being a quarter of $a_{(17)1}$ in Eq.~(\ref{A2}) also confirm this
approximation. The solution of $A^{(1)}$ for $N=24$ in Fig.~\ref{fig4}(a),
where the curve starts to oscillate,
\begin{eqnarray}
A^{(1)}|_{N=24}&=&(0.0099, 0.0153, 0.0029, \cdots -0.0030, -0.0015),
\end{eqnarray}
reveals much worse convergence, differing from Eqs.~(\ref{A1}) and (\ref{A2}) obviously,
and that the matrix elements of $U^{-1}$ have become too large.

\begin{figure}
\begin{center}
\includegraphics[scale=0.4]{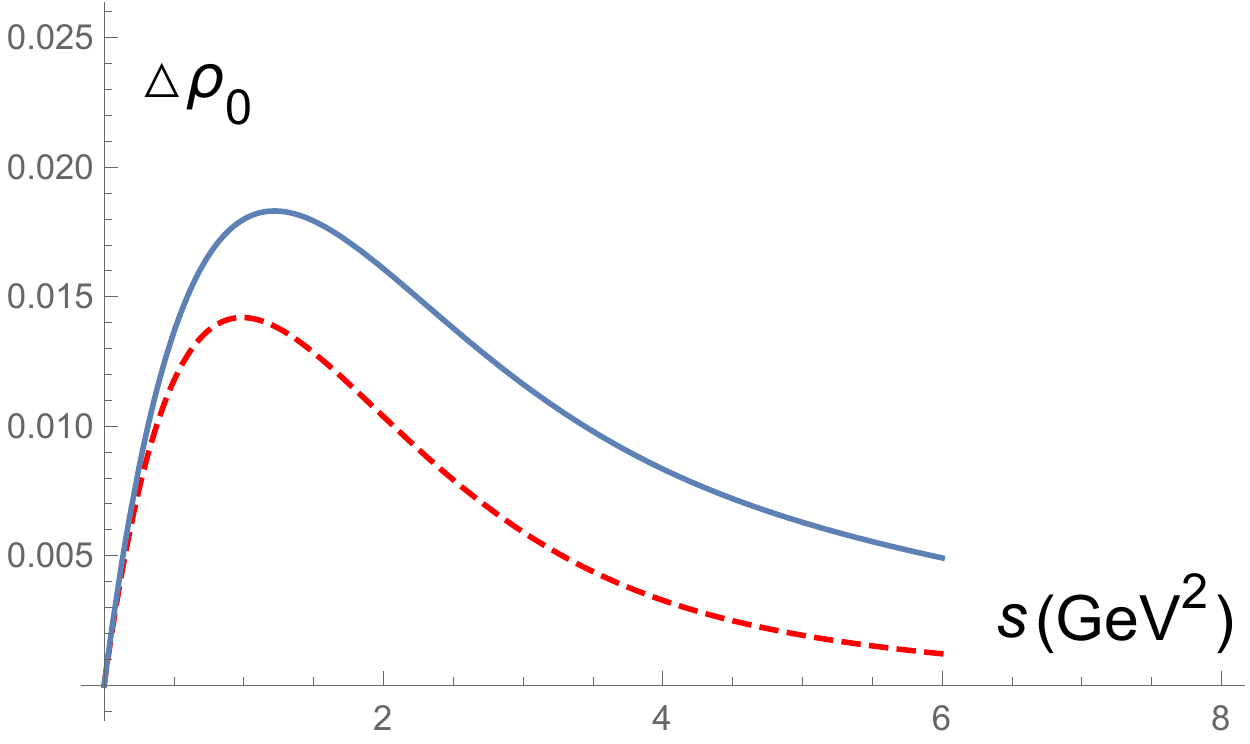}\hspace{0.5cm}
\includegraphics[scale=0.4]{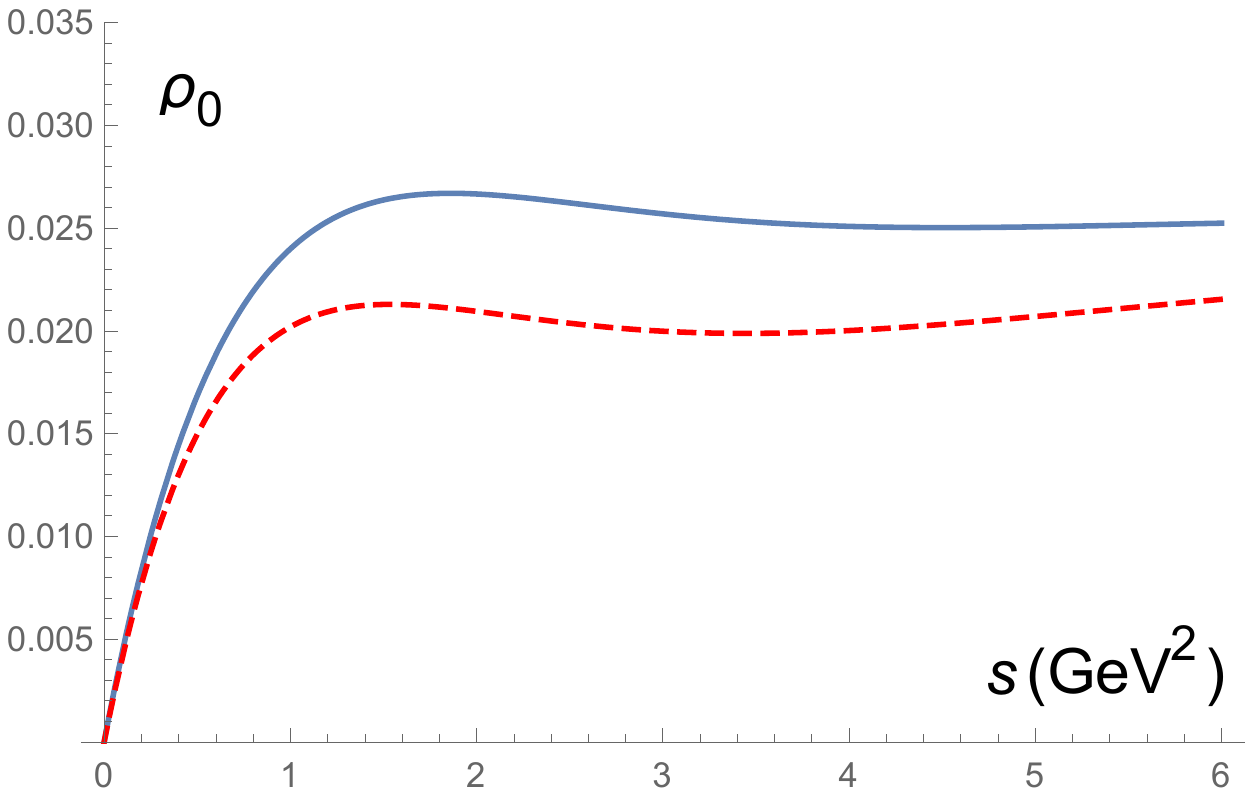}

(a) \hspace{5.0 cm} (b)
\caption{\label{fig5}
Solutions of (a) $\Delta\rho_{0}(s)$  and of (b) $\rho_{0}(s)$ for $\Lambda=2.0$ GeV$^2$, $N=10$ (dashed line)
and for $\Lambda=3.7$ GeV$^2$, $N=19$ (solid line).}
\end{center}
\end{figure}

It is easy to understand why the minima of the curves in Figs.~\ref{fig4}(a) and \ref{fig4}(b) move toward 
higher $N$, as the transition scale $\Lambda$ increases: a larger $\Lambda$ means that the region with a substantial 
continuum contribution moves toward a larger $s$. The generalized Laguerre polynomials of
higher degrees, which also take substantial values at larger $s$, are thus suitable for the expansion in 
Eq.~(\ref{r0}). This feature is explicit in Fig.~\ref{fig5}(a), where the subtracted continuum 
functions $\Delta\rho_0(s)$ for $\Lambda=2.0$ GeV$^2$ and 3.7 GeV$^2$ are exhibited. These functions are
constructed by substituting the elements $a_{jn}$ in the corresponding solutions of $A^{(1)}$ in Eqs.~(\ref{A1})
and (\ref{A2}) into Eq.~(\ref{r0}). It is seen that the taller peak of the latter is located at 
a higher $s$ relative to the shorter peak of the former. To get a complete picture, we display the
continuum functions $\rho_0(s)$ for $\Lambda=2.0$ GeV$^2$ and 3.7 GeV$^2$ in Fig.~\ref{fig5}(b), which are
obtained by adding back the subtracted pieces in Eq.~(\ref{sub}). We have verified that the curves indeed 
approach to the constant perturbative contribution ${\rm Im}I_n^{\rm pert(1)}=1/(4\pi^2)\approx 0.025$ as $s\to\infty$
(another piece from ${\rm Im}I_n^{\rm pert(2)}$ is smaller in this case). The 
shape of the continuum functions implies that the quark-hadron duality assumed in the parametrization for
the spectral density in conventional sum rules does not hold exactly, and the region
below the threshold $s_\pi=1.05$ GeV$^2$ claimed in \cite{Zhong:2021epq} still gives a
sizable contribution in fact.

It is encouraging that the minima in Figs.~\ref{fig4}(a) and \ref{fig4}(b) are both around 0.72, insensitive 
to the variation of the transition scale $\Lambda$ in a finite range. When $\Lambda$ goes above 3.7 GeV$^2$, 
the minimum disappears: the curve descends monotonically till it becomes oscillatory. We regard this 
situation as nonexistence of a solution, and the search for a stable solution stops at this maximally allowed
$\Lambda=3.7$ GeV$^2$. The solution of $\langle\xi^0\rangle|_{\mu = 2\;{\rm GeV}}=0.72$, distinct from 
unity, recalls the statement that the currently available OPE inputs are not complete. We vary the scale 
$\mu$ in Eq.~(\ref{opei3}), considering the evolution of the condensates as well, and observe that 
$\langle\xi^0\rangle$ increases as $\mu$ decreases, but cannot reach unity: even 
when $\mu$ is as low as 0.5 GeV, $\langle\xi^0\rangle$ is lifted only to 0.78 (corresponding to the maximally 
allowed $\Lambda=4.7$ GeV$^2$). As a test, we naively decrease the fourth component $B^{(1)}_4$ of the 
input vector $B^{(1)}$ by $0.01\%$, and find that the resultant $\langle\xi^0\rangle$ can be enhanced 
effectively. In other words, if the dimension-eight condensate provides a little amount of destructive correction, 
the normalization of the pion LCDA may be restored. The correction of this order of magnitude is reasonable, 
viewing that the contribution of the dimension-six condensate is about $0.1\%$ of the corresponding 
perturbative one in Eq.~(\ref{opei3}) for $\Lambda= 3.7$ GeV$^2$.  It is thus worthwhile to study the power 
corrections from the dimension-eight condensates and to examine whether the normalization of the pion LCDA is 
respected in our framework. Below we will calculate the other moments following the prescription in Eq.~(\ref{so}).

\begin{figure}
\begin{center}
\includegraphics[scale=0.4]{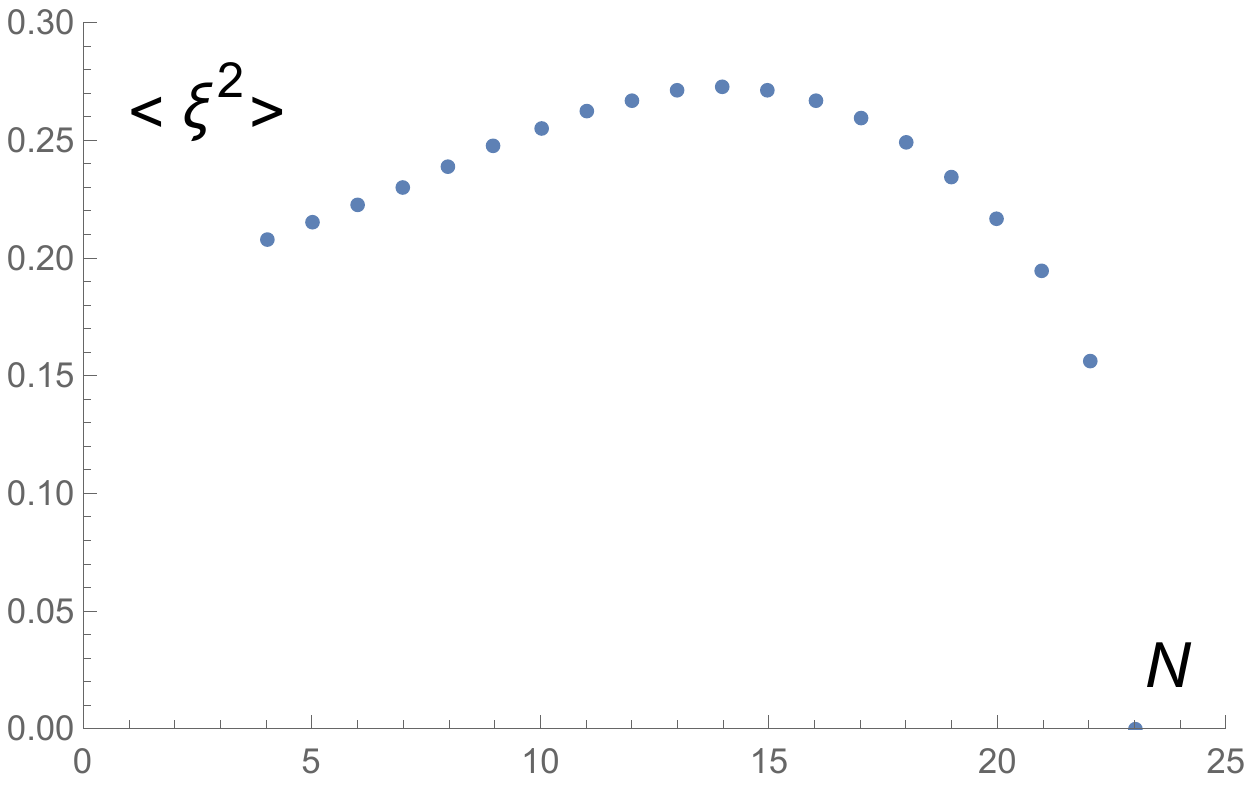}\hspace{0.5cm}
\includegraphics[scale=0.4]{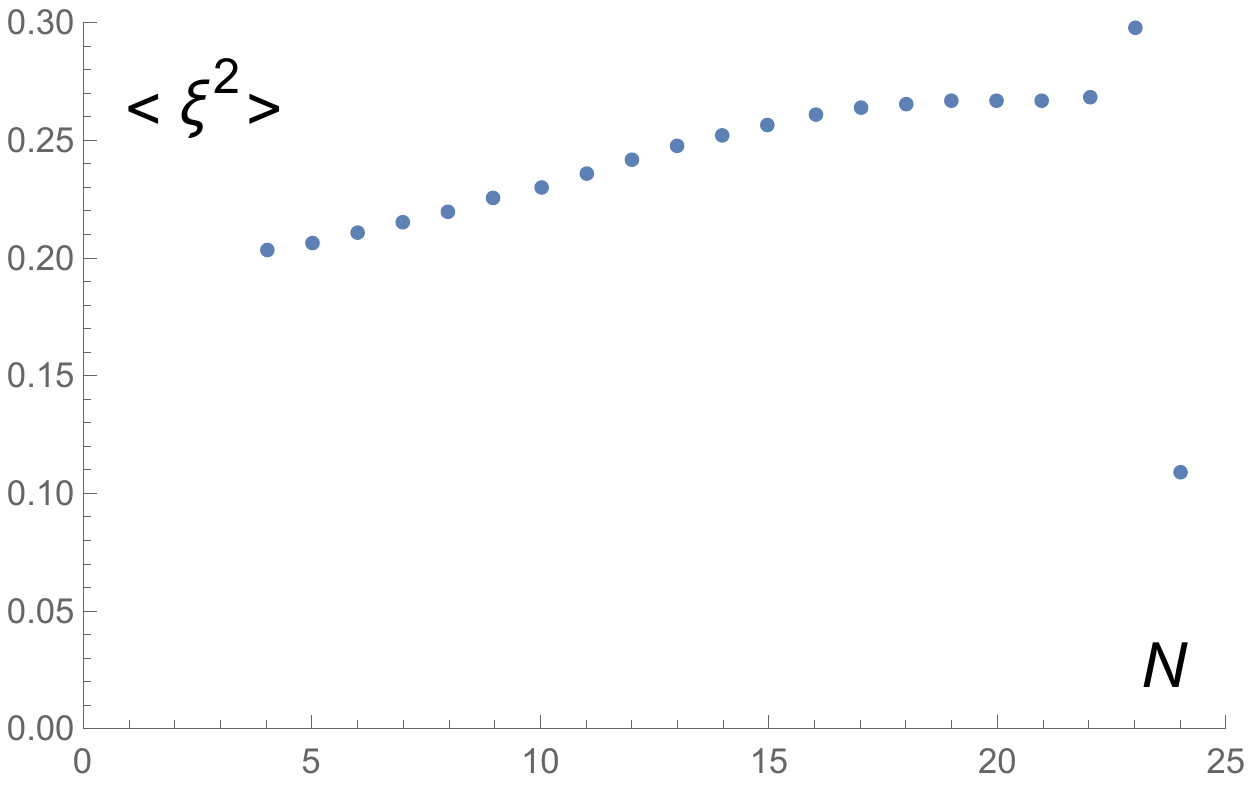}\hspace{0.5cm}
\includegraphics[scale=0.4]{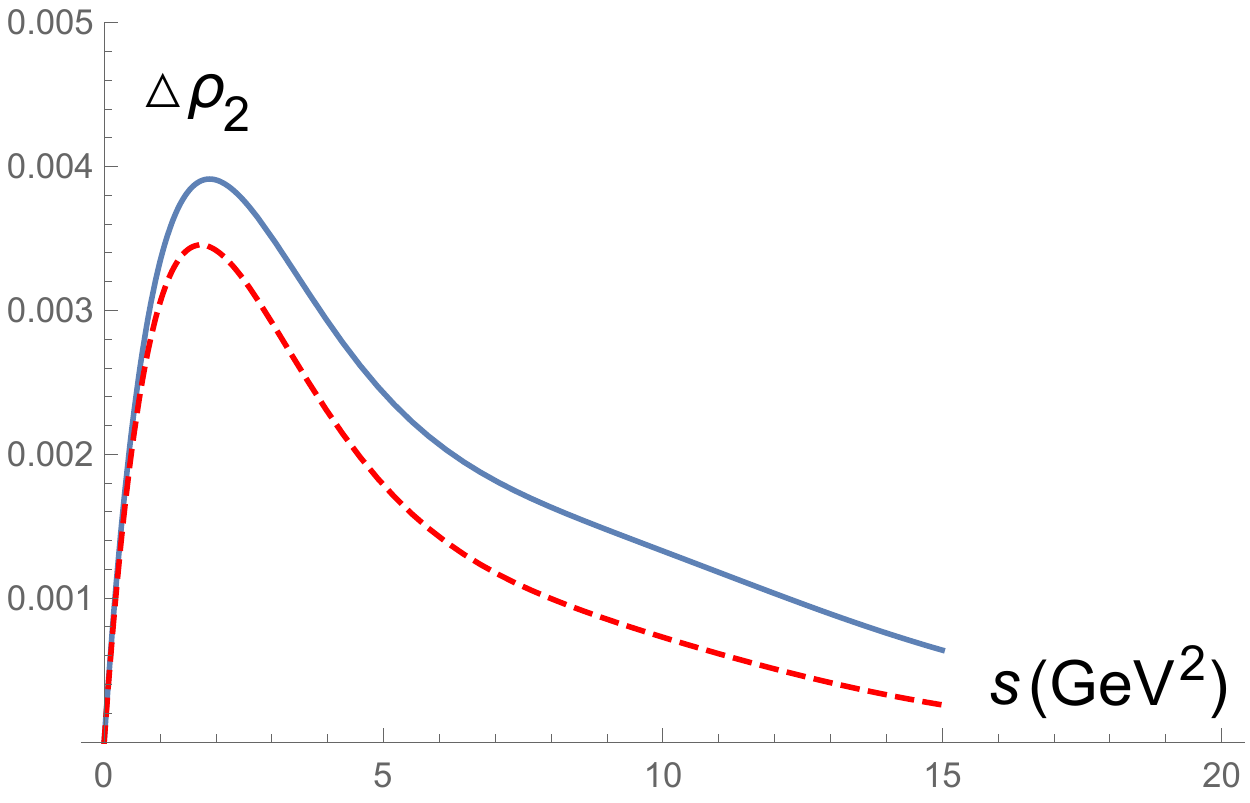}

(a) \hspace{5.0 cm} (b)\hspace{5.0 cm} (c)
\caption{\label{fig6}
Dependencies of $\langle\xi^2\rangle$ on $N$ for (a) $\Lambda=5.0$ GeV$^2$ and (b)  $\Lambda=7.2$ GeV$^2$.
(c) Solutions of $\Delta\rho_{2}(s)$ for $\Lambda=5.0$ GeV$^2$, $N=14$ (dashed line)
and for $\Lambda=7.2$ GeV$^2$, $N=20$ (solid line). }
\end{center}
\end{figure}

We repeat the above procedure to search for stable solutions of the second moment $\langle\xi^2\rangle$,
deriving the solutions of $A^{(1)}$ and $A^{(2)}$ for the same given $\Lambda$ and $N$, and taking 
the ratio of the elements $A^{(2)}_1/A^{(1)}_1$ in Eq.~(\ref{so}) to get 
$\langle\xi^2\rangle$, which is independent of the pion decay constant $f_\pi$ as mentioned before. The 
typical curves describing the $N$ dependence for $\Lambda=5$ GeV$^2$ and 7.2 GeV$^2$, which ascend with 
$N$ first, reach maxima at $N=14$ and $N=20$, and drop quickly at large $N$ as shown in Figs.~\ref{fig6}(a) 
and \ref{fig6}(b), respectively. This behavior, opposite to that in Figs.~\ref{fig4}(a) and \ref{fig4}(b), 
is attributed to the ratio of the moments that we are computing. The range in which $\langle\xi^2\rangle$ 
stays close to its maximum also broadens with $\Lambda$. In particular, the curve in Fig.~\ref{fig6}(b) 
becomes very flat around $N=20$, such that it is hard to tell the existence of a maximum. A curve for 
$\Lambda$ above 7.2 GeV$^2$ ascends monotonically, and no maximum, i.e., no solution is identified. Since the 
solutions of $\langle\xi^2\rangle$ exist at $\Lambda$ higher than for $\langle\xi^0\rangle$ in 
Fig.~\ref{fig4}, the continuum contribution associated with the former approaches to the perturbative one 
at larger $s$ according to Eq.~(\ref{sub}). This tendency is apparent in Fig.~\ref{fig6}(c), where 
the curves for the obtained subtracted continuum function $\Delta\rho_2(s)$ extend to the larger $s$ region 
compared to $\Delta\rho_0(s)$ in Fig.~\ref{fig5}(a). The peaks of $\Delta\rho_2(s)$ for $\Lambda=5$ GeV$^2$ 
and 7.2 GeV$^2$ in Fig.~\ref{fig6}(c) are located at about $s\approx 2$ GeV$^2$, corresponding to a higher 
threshold $s_\pi$, while the peaks of $\Delta\rho_0(s)$ in Fig.~\ref{fig5}(a) are located at around 1 GeV$^2$. 
That is, the thresholds $s_\pi$ are in fact different for different moments. However, the same $s_\pi$ has 
been employed in the sum-rule evaluations of the moments \cite{Zhong:2021epq}. The above features can be understood
from the viewpoint of conventional sum rules: the increase of $s_\pi$ with $\Lambda$ enhances the perturbative 
contribution, i.e., the first term on the right-hand side of Eq.~(\ref{qsr}); a larger $\Lambda$ also reduces 
the condensate contributions, which grow with $n$, an effect similar to the increase of the Borel mass $M$. 
That is, a larger $\Lambda$ facilitates the existence of a solution by improving the balance between 
the perturbative and condensate contributions. At last, it is natural that the height of the peaks in 
Fig.~\ref{fig6}(c) is a bit lower than the corresponding perturbative contribution $1/(20\pi^2)\approx 0.005$ 
due to the subtraction terms in Eq.~(\ref{sub}).

\begin{figure}
\begin{center}
\includegraphics[scale=0.5]{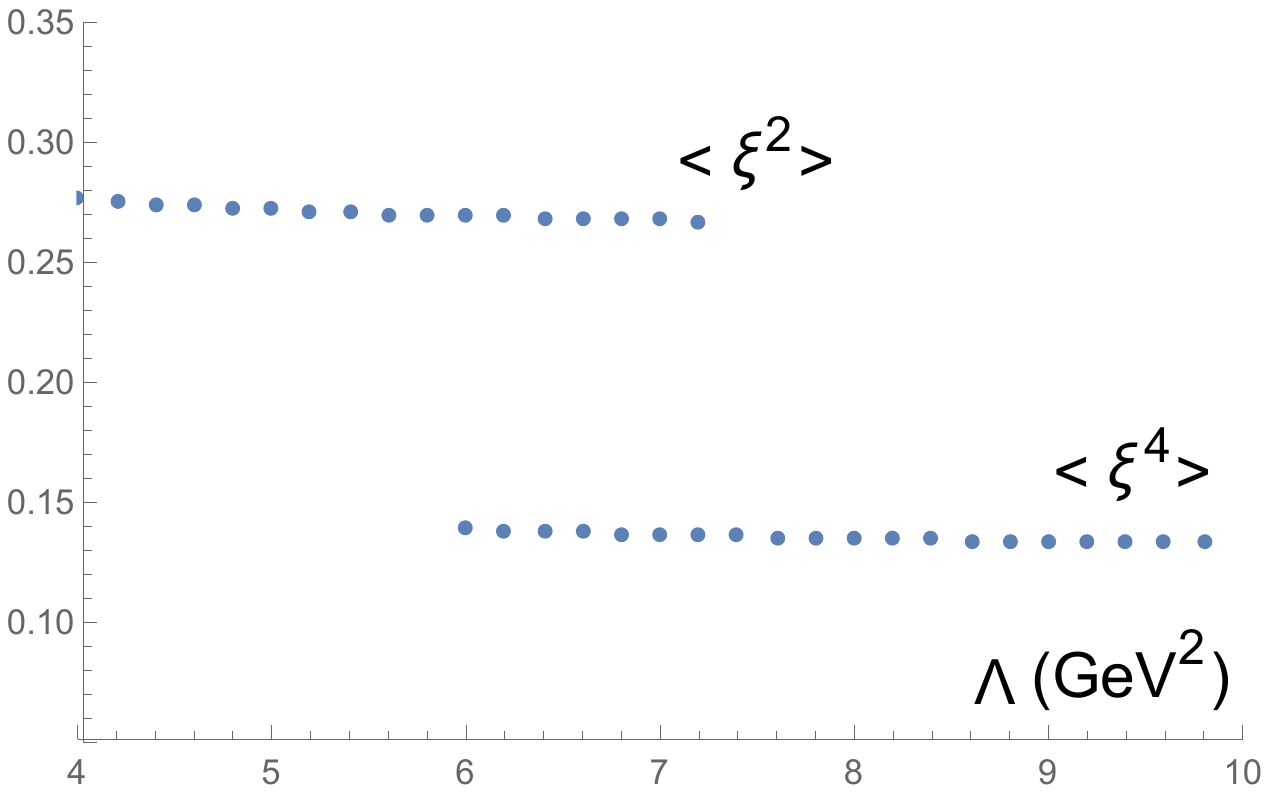}

\caption{\label{fig7}
Dependencies of $\langle\xi^2\rangle$ and $\langle\xi^4\rangle$ on $\Lambda$. }
\end{center}
\end{figure}

We read off the values of the second moment $\langle\xi^2\rangle$ from the most stable, i.e, the
best convergent solutions in $N$, like the one with $N=20$ in Fig.~\ref{fig6}(b), for various transition 
scales $\Lambda$, and plot its dependence on $\Lambda$ in Fig.~\ref{fig7}. It is seen that the curve is 
quite stable with respect to the change of $\Lambda$. We select the value corresponding to the 
maximally allowed $\Lambda$ as our result for $\langle\xi^2\rangle$. Other points on the curve are also 
acceptable solutions, whose values differ from the selected one by about 3\%, which reflects the stability 
of the solutions. The same procedure is then applied to the evaluation of higher moments, and the resultant dependence
of the fourth moment $\langle\xi^4\rangle$ is also plotted in Fig.~\ref{fig7}, which reveals the similar 
behavior with excellent stability. In principle, we can obtain all the moments of the pion LCDA
in this manner, but list only the first few below as examples,
\begin{eqnarray}
(\langle\xi^2\rangle, \langle\xi^4\rangle,\langle\xi^6\rangle,\langle\xi^8\rangle,\langle\xi^{10}\rangle,
\langle\xi^{12}\rangle,\cdots)|_{\mu = 2\,{\rm GeV}}=(0.2672, 0.1333,0.0871,0.0658, 0.0546, 0.0480,\cdots).
\label{pm4}
\end{eqnarray}
The above values are a bit higher than, for instance, $\langle\xi^2\rangle|_{\mu = 2\;{\rm GeV}}=0.254$, 
in \cite{Zhong:2021epq}. However, we remind that the input of the triple-gluon condensate 
has been modified into Eq.~(\ref{gg3}), which differs from the one in \cite{Zhong:2021epq}.
Our $\langle\xi^2\rangle|_{\mu = 2\;{\rm GeV}}$ is also larger than in other methods (see Table~III in
\cite{Zhong:2021epq} for a summary and more complete references), such as $0.210\pm 0.013\;{\rm (stat.)} 
\pm 0.034\;{\rm (sys.)}$ from the recent lattice QCD analysis performed at a pion mass $m_\pi=550$ MeV 
\cite{Detmold:2021qln}.

We have examined the sensitivity of the solutions to the OPE inputs, and found that the condensates
$\langle m_q\bar q q\rangle$ and $m_q \langle g_s\bar{q}\sigma TGq\rangle$ have little influence.
For example, multiplying the latter by a factor of 2 has no effect at all. Adopting the lower (upper) 
bound the gluon condensate $\langle\alpha_s G^2\rangle$ in Eq.~(\ref{mqq}), we have 
$\langle\xi^2\rangle|_{\mu = 2\;{\rm GeV}}=0.2860$ (0.2582) at $N=20$ for the maximally allowed
$\Lambda=6.2$ GeV$^2$ ($\Lambda=8.2$ GeV$^2$). It indicates that the $\pm 29\%$ change of 
$\langle\alpha_s G^2\rangle$ causes only $^{-3}_{+7}\%$ uncertainty of the second moment 
$\langle\xi^2\rangle|_{\mu = 2\;{\rm GeV}}$, and that our results are less sensitive to
the variation of the dimension-four condensates. The dimension-six condensates provide the major source 
of theoretical uncertainties, which can be illustrated by varying the triple-gluon condensate 
$\langle g_s^3fG^3\rangle$: the lower (upper) bound in Eq.~(\ref{gg3}) leads to
$\langle\xi^2\rangle|_{\mu = 2\;{\rm GeV}}=0.2860$ (0.2529) at $N=20$ for the maximally allowed
$\Lambda=6.8$ GeV$^2$ ($\Lambda=8.0$ GeV$^2$). That is, the $\pm 12\%$ change of the triple-gluon condensate 
in Eq.~(\ref{gg3}) yields $^{-5}_{+7}\%$ error. The above investigation explores the theoretical
uncertainties of our results attributed to the variation of the condensate inputs, which is of order of 10\%.

Converting the moments in Eq.~(\ref{pm4}), which seem to converge satisfactorily, into the Gegenbauer coefficients 
via Eq.~(\ref{cv}), we get 
\begin{eqnarray}
\label{p1}
(a_2^\pi,a_4^\pi,a_6^\pi, a_8^\pi, a_{10}^\pi,a_{12}^\pi,\cdots)|_{\mu=2\,{\rm GeV}}=
(0.1960, 0.0268, 0.1918, 0.1376, 0.4034,  -0.1319,\cdots),\label{pga1}
\end{eqnarray} 
with worse convergence. The $x$ dependence of the pion LCDA constructed from the above Gegenbauer coefficients  
oscillates between the values $-1$ and 3, more violently than in Fig.~\ref{fig1}. It is unlikely to be a reasonable form 
of a LCDA, manifesting the difficulty to acquire the $x$ dependence of the pion LCDA from the moments, even when all 
the moments are known. Therefore, we turn to the direct extraction of  the pion LCDA 
from the dispersion relations for the Gegenbauer coefficients.

\subsection{$x$ Dependence of the pion LCDA}

\begin{figure}
\begin{center}
\includegraphics[scale=0.4]{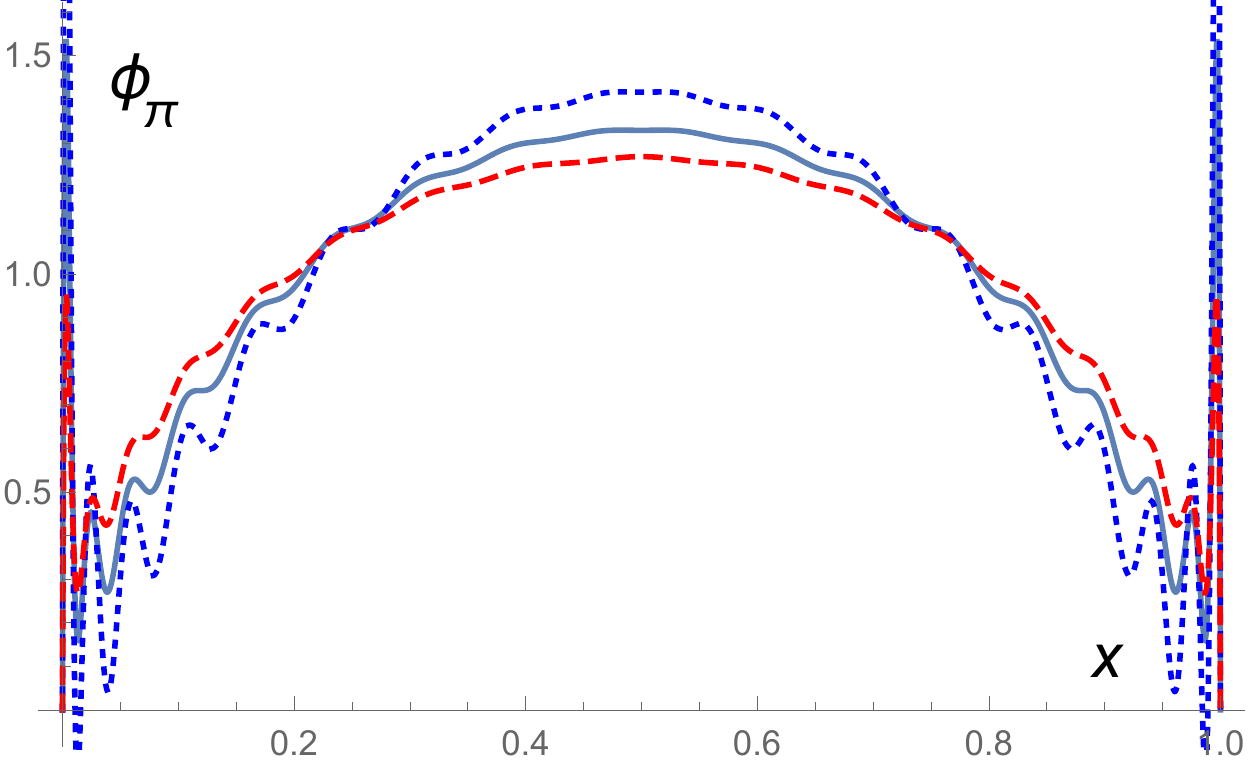}\hspace{0.5cm}
\includegraphics[scale=0.4]{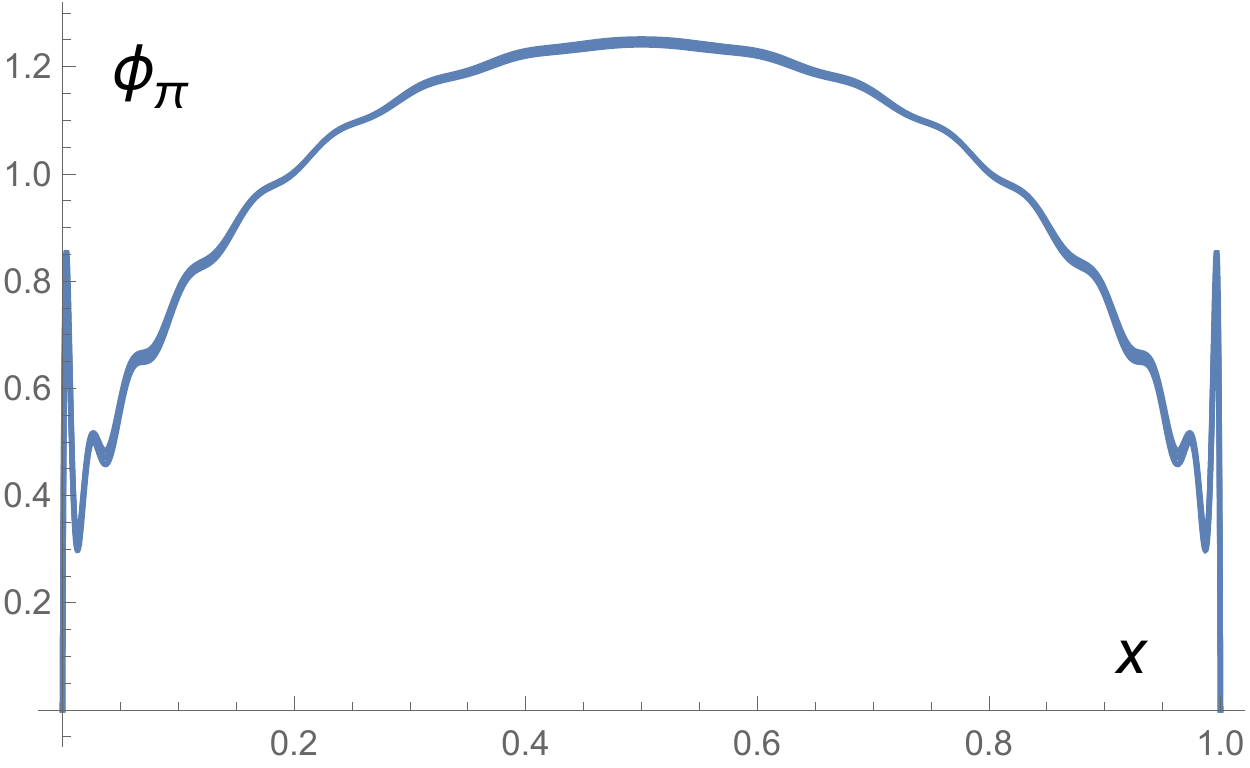}\hspace{0.5cm}
\includegraphics[scale=0.4]{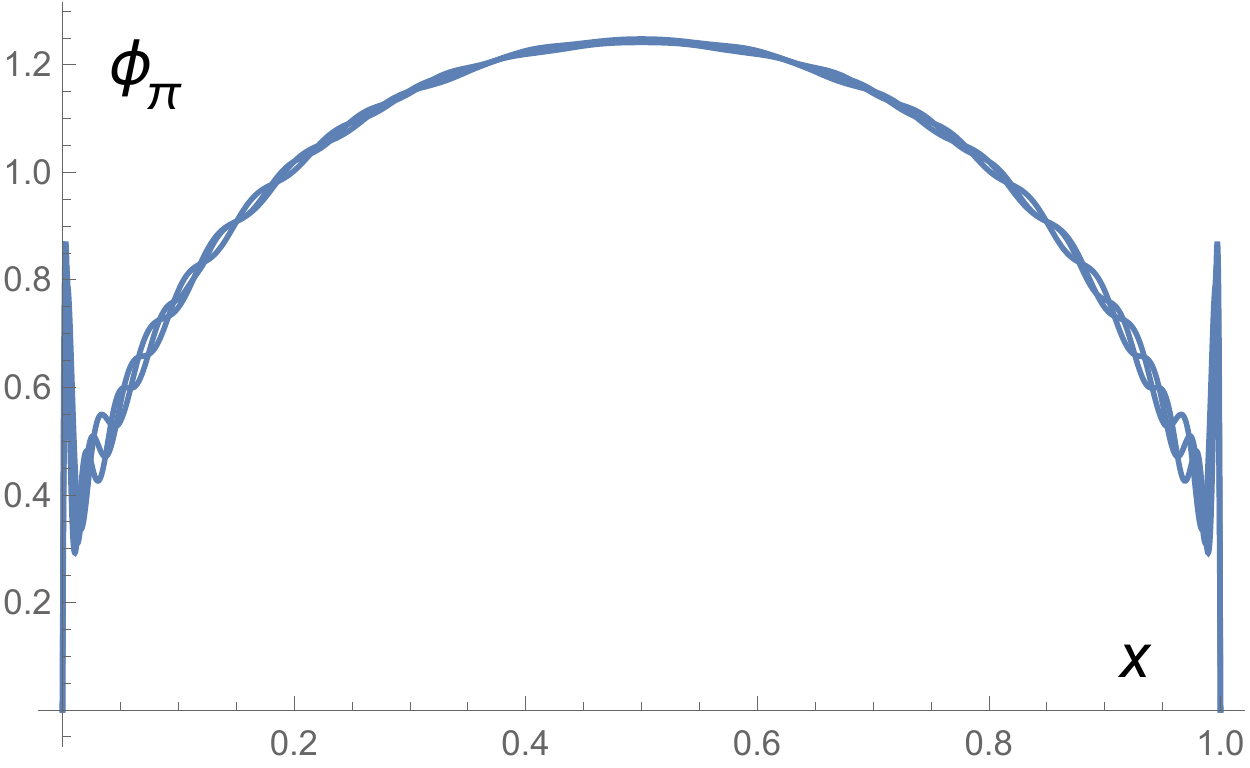}

(a) \hspace{5.0 cm} (b)\hspace{5.0 cm} (c)
\caption{\label{fig8}
Dependencies of the pion LCDA $\phi_\pi(x)$ for (a)  $\lambda=0.01$ ($\Lambda=13.01$ GeV$^2$,
dotted line), 0.1 ($\Lambda=12.37$ GeV$^2$, solid line) and 
$0.3$ ($\Lambda=12.07$ GeV$^2$, dashed line) with $N=18$, (b) $\lambda=0.40$ ($\Lambda=12.01$ GeV$^2$), 
0.45 ($\Lambda=11.99$ GeV$^2$) and $0.50$ ($\Lambda=11.97$ GeV$^2$) with $N=18$, and
(c) $\lambda=0.45$ with $N=16$ ($\Lambda=10.28$ GeV$^2$), 18 ($\Lambda=11.99$ GeV$^2$), 
and 20 ($\Lambda=13.72$ GeV$^2$).}
\end{center}
\end{figure}

We apply the formalism developed in Sec.~IIC to the determination of the Gegenbauer coefficients for
the pion LCDA from the OPE inputs. The construction of the $N\times N$ matrices $U$ and $V$ follow Eqs.~(\ref{mm2}) 
and (\ref{vm}), respectively, and the elements $B_{in}$ of the input matrix $B$ are computed from
Eqs.~(\ref{opei1})-(\ref{opei}). The solutions $A=U^{-1}B(V+\lambda I)^{-1}$ are then obtained straightforwardly, 
whose first rows give the Gegenbauer coefficients. We increase the regularization parameter $\lambda$ 
step by step, and search for stable solutions in the way similar to that in the previous subsection. 
It is immediately noticed that no stable solutions exist for a vanishing regularization parameter $\lambda=0$: 
the magnitudes of the Gegenbauer coefficients $a_{2n-2}^\pi$ always grow fast with $n$, no matter how the 
transition scale $\Lambda$ and the dimension $N$ are tuned, leading to a shape of the pion LCDA as 
in Fig.~\ref{fig3}(a). This is not a surprise, because the OPE inputs are incomplete and not 
accurate enough, such that solutions from the inverse matrix method are not well tamed. It turns out that 
$\lambda$ has to be sizable in order to effectively suppress the divergent behavior, and to allow solutions
of $A$, which are convergent under the variations of $N$ and $\Lambda$. The present case differs from the 
analysis on the mock data, which represent precise inputs, and require only a milder regularization. 
We then read the Gegenbauer coefficients from $A$, and use them to construct the pion LCDA in the Gegenbauer expansion.

Figure~\ref{fig8}(a) exhibits the $x$ dependencies of the pion LCDA $\phi_\pi(x)$ for the increasing $\lambda=0.01$, 
0.1, and 0.3, which correspond to a sufficiently high dimension $N=18$ and the transition scales $\Lambda=13.01$ 
GeV$^2$, 12.37 GeV$^2$, and 12.07 GeV$^2$, respectively. It is found that the solutions stabilize with $\lambda$ 
steadily: the curve for $\lambda=0.01$ is oscillatory, while the curve for $\lambda=0.3$ becomes relatively smooth. 
We further increase $\lambda$, and observe that the shapes of the pion LCDA for $\lambda>0.3$ remain almost the same 
and independent of $\lambda$:  the three curves of $\phi_\pi(x)$ with $N=18$ for $\lambda=0.40$, 0.45 and 0.50, 
corresponding to $\Lambda=12.01$ GeV$^2$, $11.99$ GeV$^2$ and $11.97$ GeV$^2$, respectively, overlap perfectly as 
shown in Fig.~\ref{fig8}(b). The spikes near the endpoints of $x$ would be pushed toward $x=0$ and 1 and disappear 
from the considered domain, if the dimension $N$ could continue to increase. Other than the endpoint behavior, the 
shape of $\phi_\pi(x)$ in Fig.~\ref{fig8}(b) is smooth. We also display $\phi_\pi(x)$ with $\lambda=0.45$ for 
$N=16$, 18 and 20, corresponding to $\Lambda=10.28$ GeV$^2$, 11.99 GeV$^2$ and 13.72 GeV$^2$, respectively, in 
Fig.~\ref{fig8}(c). The overlap of the three curves is also remarkable, implying that the pion LCDA resulting from 
the summation of the contributions up to 18 Gegenbaer polynomials is already stable with the variations of $N$ and 
$\Lambda$. The bands in both Figs.~\ref{fig8}(b) and (c) reflect the theoretical uncertainty of the pion LCDA in 
the inverse matrix method. In terms of the Gegenbauer coefficient $a_2^\pi$, we have $a_2^\pi=0.1735$, 0.1775 and 
0.1811 with $N=18$ for $\lambda=0.40$, 0.45 and 0.50, respectively, and $a_2^\pi=0.1814$, 0.1775 and 0.1748 for 
$\lambda=0.45$ with $N=16$, 18 and 20, respectively. Namely, the finite stability ranges in $\lambda$ and in $N$ 
give about 2\% errors to the obtained Gegenbauer coefficients.

\begin{table}[tbh]
	\centering
	\caption{Comparison of $a_2^\pi$ and $a_4^\pi$ for the pion LCDA at $\mu=2$ GeV in different methods,
	where the numbers without (with) parentheses in LFQM \cite{Choi:2007yu} were derived from the linear 
	(harmonic oscillator) potential.
	The results presented at $\mu=1$ GeV in \cite{Choi:2007yu,Mikhailov:2021znq,Cheng:2020vwr,Hua:2020usv} have 
	been evolved to $\mu=2$ GeV.}
	\begin{tabular}{lcc}
		\hline
		Methods&$a_2^\pi$&$a_4^\pi$\\
		\hline
		This work&$0.1775^{+0.0036}_{-0.0040}$ &$0.0957^{+0.0011}_{-0.0012}$\\
		Lattice QCD \cite{Bali:2019dqc}&$0.101\pm 0.023$&\\
		Lattice QCD \cite{Hua:2022kcm}&$0.258\pm 0.087$&$0.122\pm 0.055$\\
		Lattice QCD \cite{Arthur:2010xf}&$0.233\pm 0.065$&\\
		Lattice QCD \cite{Braun:2015axa}&$0.136\pm 0.021$&\\
		QCD sum rules \cite{Stefanis:2014nla}&$0.057^{+0.024}_{-0.019}$&$-0.013^{+0.022}_{-0.019}$\\
        QCD sum rules \cite{BMS}&$0.149^{+0.052}_{-0.043}$&$-0.096^{+0.063}_{-0.058}$\\
        QCD sum rules \cite{Zhong:2021epq}&$0.157\pm 0.029$&$0.032\pm 0.007$\\
        LFQM \cite{Choi:2007yu}&0.092 (0.038)&-0.002 (-0.020)\\
		LCSR fit \cite{Mikhailov:2021znq}&0.085&$-0.020$\\
		LCSR fit \cite{Cheng:2020vwr}&$0.205 \pm 0.036$ &$0.125 \pm 0.042$\\
		Global fit \cite{Hua:2020usv}&$0.491 \pm 0.058$ &$0.084 \pm 0.029$\\
		\hline
	\end{tabular}\label{table}
\end{table}

We gather our solutions for the set of Gegenbauer coefficients  
\begin{eqnarray}
& &(a_2^\pi,a_4^\pi,a_6^\pi, a_8^\pi, a_{10}^\pi,a_{12}^\pi,\cdots,a_{32}^\pi,a_{34}^\pi)|_{\mu=2\,{\rm GeV}}\nonumber\\
&=&(0.1775^{+0.0036}_{-0.0040}, 0.0957^{+0.0011}_{-0.0012}, 0.0762^{+0.0006}_{-0.0003}, 
0.0688^{+0.0016}_{-0.0012}, 0.0643^{+0.0021}_{-0.0017}, 0.0603^{+0.0024}_{-0.0019},\nonumber\\
& &\cdots, 0.0089^{+0.0004}_{-0.0006}, 0.0028^{+0.0001}_{-0.0003}),
\label{ga6}
\end{eqnarray}
where the central values correspond to $\lambda=0.45$ and $N=18$, and the errors from the variation of 
$\lambda$ in the interval $[0.40,0.50]$ correspond to the band in Fig.~\ref{fig8}(b). It is apparent
that the elements in Eq.~(\ref{ga6}) are all under control in contrast to those in Eq.~(\ref{pga1}).
The convergent sequence in Eq.~(\ref{ga6}) accounts for the smoothness of the pion
LCDA in Figs.~\ref{fig8}(b) and \ref{fig8}(c). The pion LCDA with the above coefficients yields the moments
with the central values
\begin{eqnarray}
(\langle \xi^2 \rangle,\langle \xi^4 \rangle,\langle \xi^6 \rangle,
\langle \xi^8 \rangle,\langle \xi^{10} \rangle,\langle \xi^{12} \rangle,\cdots)|_{\mu = 2\,{\rm GeV}}=
(0.2609, 0.1362, 0.0890, 0.0652, 0.0511, 0.0420,\cdots).\label{momp}
\end{eqnarray}
whose agreement with Eq.~(\ref{pm4}) verifies the consistency of implementing the regularization. 
We remind that the Gegenbauer coefficients in Eq.~(\ref{ga6}) are derived at the same $\Lambda$ as an 
attempt to achieve their convergence, while each of the moments in Eq.~(\ref{pm4}) is evaluated at a 
separate $\Lambda$. This causes the minor difference between Eqs.~(\ref{pm4}) and (\ref{momp}). The comparison
of the first two Gegenbauer coefficients in Eq.~(\ref{ga6}) with the results in the literature is 
summarized in Table~\ref{table}. The value of 
$a_2^\pi$ in Eq.~(\ref{ga6}) is of the same order as those obtained in other methods, such as lattice QCD 
\cite{Arthur:2010xf,Hua:2022kcm,Braun:2015axa,Bali:2019dqc}. Our $a_4^\pi$ is too, but differs from those in QCD 
sum rules \cite{Stefanis:2014nla,BMS}, from the light-front quark model (LFQM)
\cite{Choi:2007yu}, and from the indication of the data of the pion transition form factor analyzed in
light-cone sum rules (LCSR) \cite{Mikhailov:2016klg,Stefanis:2020rnd,Mikhailov:2021znq}, which 
tend to be negative. However, the fits to the data of the pion form factors based on LCSR
favor positive $a_4^\pi$ \cite{Khodjamirian:2011ub,Cheng:2020vwr}.
Note that $a_2^\pi$ and $a_4^\pi$ in Eq.~(\ref{ga6}) are distinct from those determined in the global study of 
two-body hadronic $B$ meson decays formulated in the perturbative QCD approach \cite{Hua:2020usv}, where
the leading-order factorization formulas were employed. Hence, it is worth including subleading 
contributions to the above decays in the perturbative QCD approach, performing a global analysis with higher precision,
and checking whether fit results of the Gegenbauer coefficients become closer to those in Eq.~(\ref{ga6}).
Besides, the Gegenbauer coefficients presented here depend on the inputs for the condensates,
which are not yet completely certain.

\begin{figure}
\begin{center}
\includegraphics[scale=0.4]{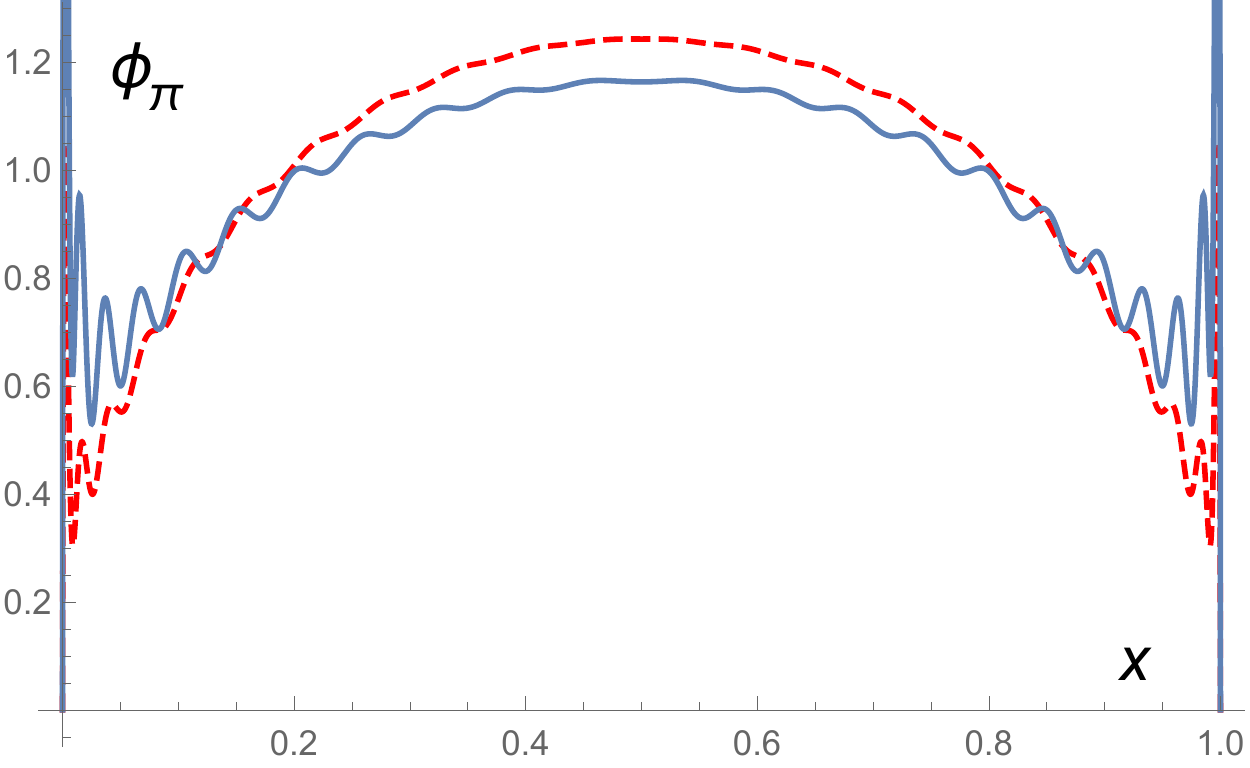}\hspace{0.5cm}
\includegraphics[scale=0.4]{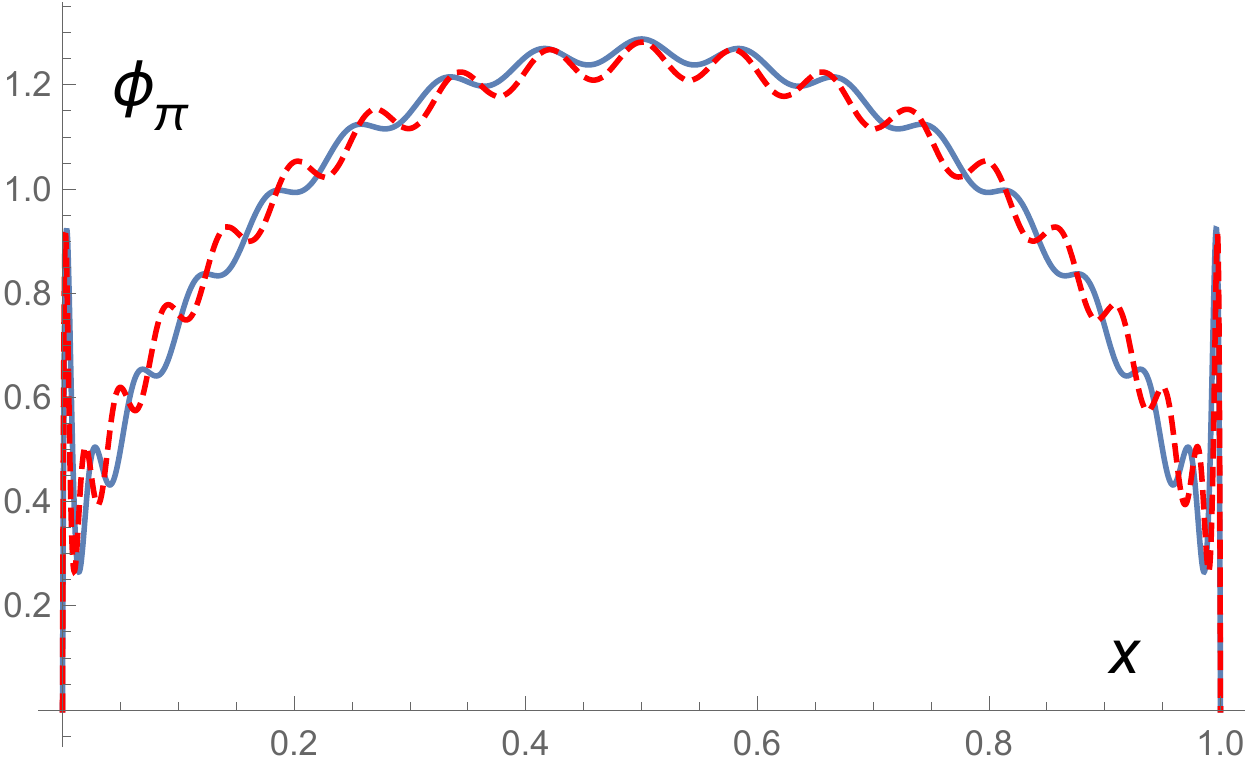}

(a) \hspace{5.0 cm} (b)
\caption{\label{fig9}
Dependencies of the pion LCDA $\phi_\pi(x)$ for (a) $\lambda=0.45$ with $N=22$
($\Lambda=15.72$ GeV$^2$, dashed line) and $N=23$  ($\Lambda=17.57$ GeV$^2$, solid line),
and (b) $\lambda=0.30$, $\Lambda=12$ GeV$^2$ and $N=18$ (solid line),
and $\lambda=0.45$, $\Lambda=13$ GeV$^2$ and $N=19$ (dashed line).}
\end{center}
\end{figure}

To get a picture of the behavior of the pion LCDA slightly away from the above best convergent solutions,
we display $\phi_\pi(x)$ for $\lambda=0.45$ with $N=22$ ($\Lambda=15.72$ GeV$^2$) and $N=23$  ($\Lambda=17.57$ GeV$^2$) 
in Fig.~\ref{fig9}(a). The former represents a solution outside the stability region, and the latter represents a 
solution starting to go out of control with larger $N$ and $\Lambda$. For $N=22$, the shape of the curve is still 
similar to that for $N=16$-20 in Fig.~\ref{fig8}(c), except that it is bumpier due to the worse convergence of the 
Gegenbauer coefficients, and the spikes near the endpoints are sharper and squeezed further to the endpoints. As the 
dimension increases to $N=23$, the shape becomes more oscillatory with significant spikes, and also deviates more 
from that for $N=16$-20. Figure~\ref{fig9}(b) collects two results of $\phi_\pi(x)$ with undesired behaviors: the 
solid line corresponds to another transition scale $\Lambda=12$ GeV$^2$ for $\lambda=0.3$ and $N=18$, 
which differs from the solution with $\Lambda=12.07$ GeV$^2$ for the same $\lambda$ and $N$ in Fig.~\ref{fig8}(a).
The dashed line arises from randomly chosen $\Lambda=13$ GeV$^2$ and $N=19$ for $\lambda=0.45$.
In both cases the bad convergence of the Gegenbauer coefficients induces numerous oscillations
due to the ill-posed nature, though the overall shapes of the curves remain similar.
The comparison between Figs.~\ref{fig8}(b), \ref{fig8}(c) and \ref{fig9} indicates that the existence of the smooth
solutions for the pion LCDA is nontrivial.

Fitting the parametrization for the pion LCDA
\begin{eqnarray}
\frac{\Gamma(2p+2)}{\Gamma(p+1)^2}x^p(1-x)^p, \label{pp}
\end{eqnarray}
which has been normalized appropriately, to the curves in Fig.~\ref{fig8}(b), we deduce
\begin{eqnarray}
p=0.45\pm 0.02,\label{ppv}
\end{eqnarray}
where the upper (lower) bound of the error comes from the regularization parameter $\lambda=0.5$ (0.4). 
The leading-twist pion LCDA has been analyzed in various approaches, and the results were summarized briefly in 
\cite{Zhong:2021epq,Hua:2022kcm}. It is interesting to find that our LCDA described by Eq.~(\ref{ppv}) is 
very close to the one from the dynamical-chiral-symmetry-breaking-improved kernel for the Dyson-Schwinger equations 
\cite{Chang:2013pq,Roberts:2021nhw}, and to the ones from the random-instanton-vacuum model \cite{Kock:2021spt},
the Nambu-Jona-Lasinio model \cite{Broniowski:2017wbr}, the nonlocal chiral quark model \cite{Nam:2017gzm},
the light-cone quark model \cite{Kaur:2020vkq} and the basis light front quantization \cite{Mondal:2021czk}. 
As mentioned before, results from the Dyson-Schwinger equations
depend on kernels: the rainbow-ladder kernel \cite{Chang:2013pq} leads to a shorter and broader pion LCDA with a 
shape different from that in Fig.~\ref{fig8}(b). On the other hand, our LCDA is a bit narrower than that from the 
recent lattice QCD calculation based on the quasi-light-front correlation function \cite{Hua:2022kcm}.

\begin{figure}
\begin{center}
\includegraphics[scale=0.4]{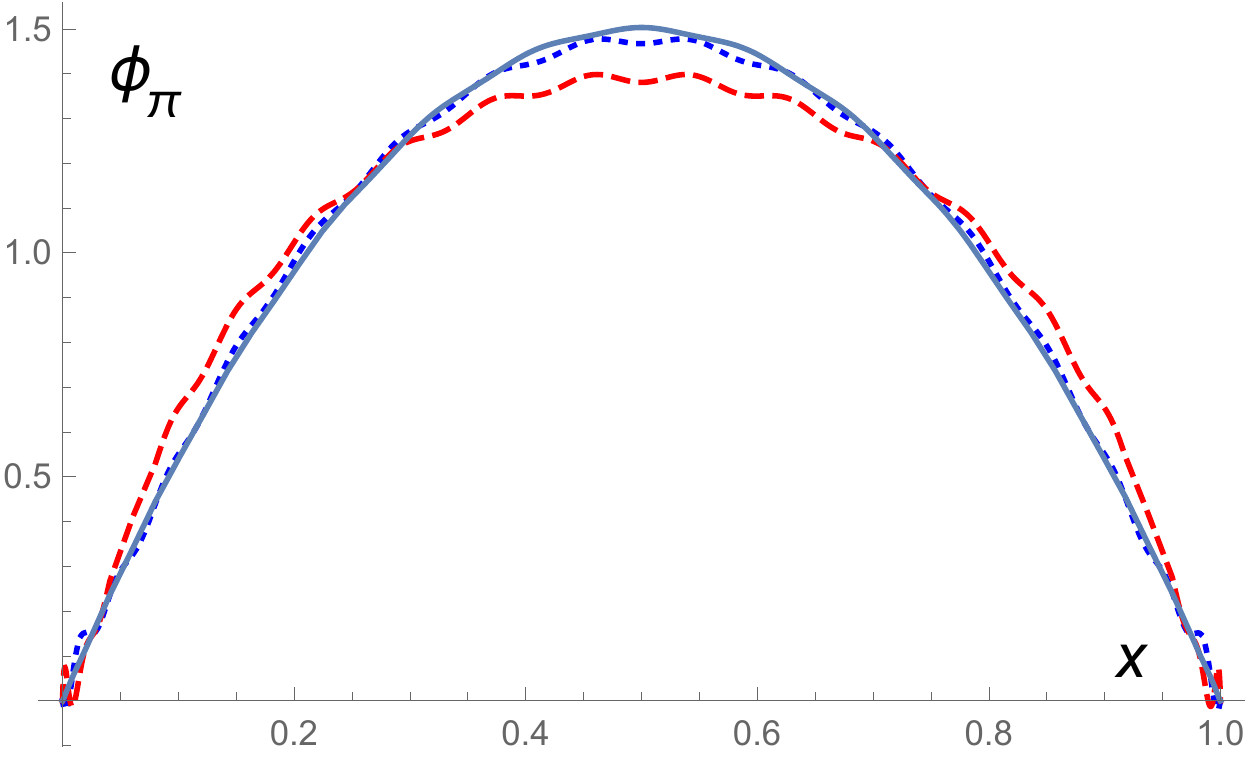}\hspace{0.5cm}
\includegraphics[scale=0.4]{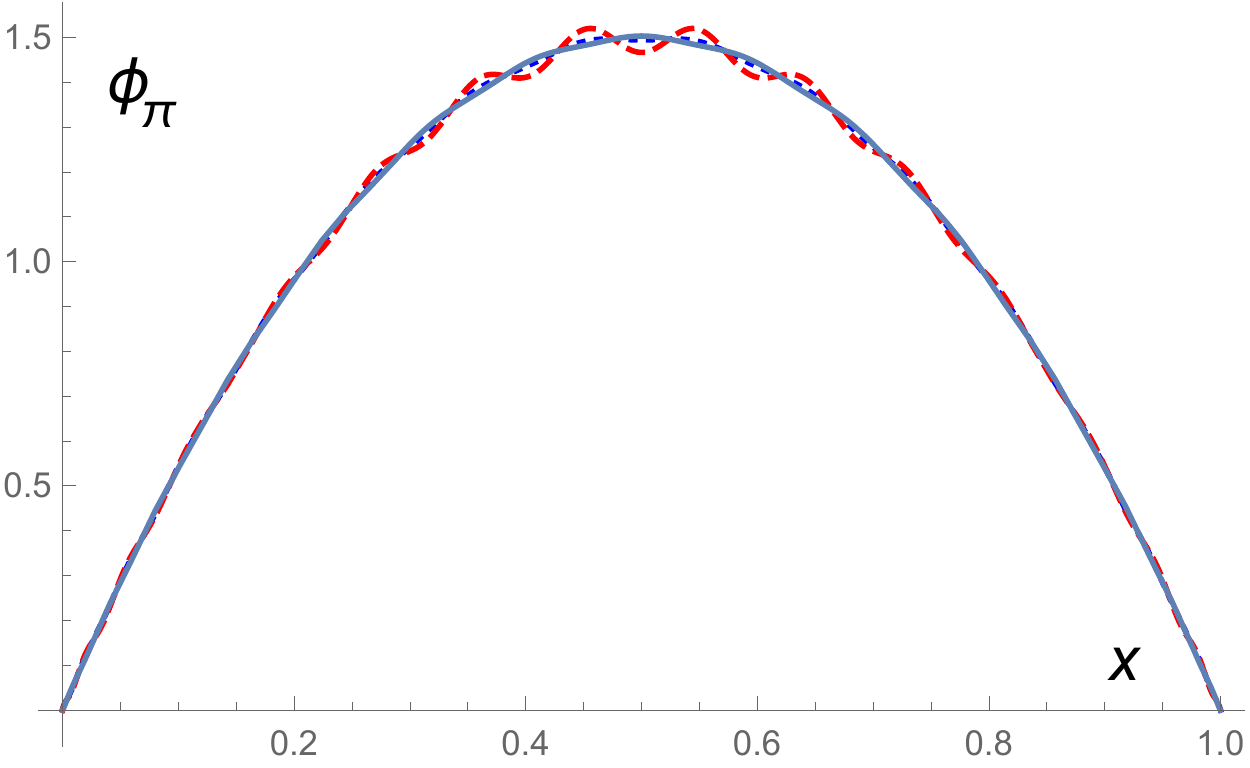}

(a) \hspace{5.0 cm} (b)
\caption{\label{fig10}
Dependencies of the pion LCDA $\phi_\pi(x)$ for (a) $\lambda=0$ (solide line), 0.01 (dotted line)
and $0.10$ (dashed line) with $N=18$, and (b) $\lambda=0$ with $N=16$ (dotted line), 18 (solid line) 
and 20 (dashed line) in the absence of the condensates.}
\end{center}
\end{figure}

As a check, we seek the solutions of the pion LCDA $\phi_\pi(x)$ in the sole presence of the perturbative 
piece in the OPE. It is easily seen that solutions are independent of the scale $\mu$ without the 
condensates at the considered level of accuracy. Moreover, they are almost independent of the transition scale 
$\Lambda$, because its dependence appears only through the tiny ratio $r_m=m_\pi^2/\Lambda$. Similarly, 
we increase the regularization parameter $\lambda$ gradually, and examine the stability of the 
solutions. It is observed that solutions are insensitive to the variation 
of $\lambda$ till it reaches about 0.1. The $x$ dependencies of the pion LCDA for $\lambda=0$, 0.01 and 0.1 with 
the dimension $N=18$ are exhibited in Fig.~\ref{fig10}(a), where the curves for $\lambda=0$ and 0.01 overlap 
well, demonstrating the stability of the solutions. No spikes near the endpoints of $x$ show up, since the OPE 
inputs are simple in this case. As $\lambda$ increases to 0.1, the shape of the pion LCDA becomes bumpy with a 
shorter height in the intermediate $x$ region due to worse convergence in $N$. 
We also present the $x$ dependencies of the pion LCDA for $N=16$, 18, and 20
with $\lambda=0$ in Fig.~\ref{fig10}(b), where the curves for $N=16$ and 18 overlap perfectly
without visible difference. The curve for $N=20$ is bumpy in the intermediate $x$ region,
though the shape remains the same, implying that the ill-posed nature starts to appear.
The set of Gegenbauer coefficients corresponding to $\lambda=0$ and $N=18$ is given by
\begin{eqnarray}
& &(a_2^\pi,a_4^\pi,a_6^\pi, a_8^\pi, a_{10}^\pi,a_{12}^\pi,\cdots,a_{32}^\pi,a_{34}^\pi)\nonumber\\
&=&(\sim 0, \sim 0, \sim 0, \sim 0, \sim 0, 4.4\times 10^{-9},\cdots, 0.0001, -0.0001),
\label{ga7}
\end{eqnarray}
where $\sim 0$ denotes a value smaller than $10^{-10}$, and the last two coefficients of $O(10^{-4})$
are attributed to the growing elements of the inverse matrix $V^{-1}$. It is obvious that the above
Gegenbauer coefficients describe a pion LCDA in the asymptotic form. This result, suggesting that
the deviation of the pion LCDA from the asymptotic form is caused by the condensates,
further supports the consistency of our formalism. It also confirms that the balance
between perturbative and condensate contributions is not a crucial requirement for the existence of stable 
solutions in our approach, because the condensates are absent in the above analysis.

\subsection{QCD Evolution}

\begin{figure}
\begin{center}
\includegraphics[scale=0.4]{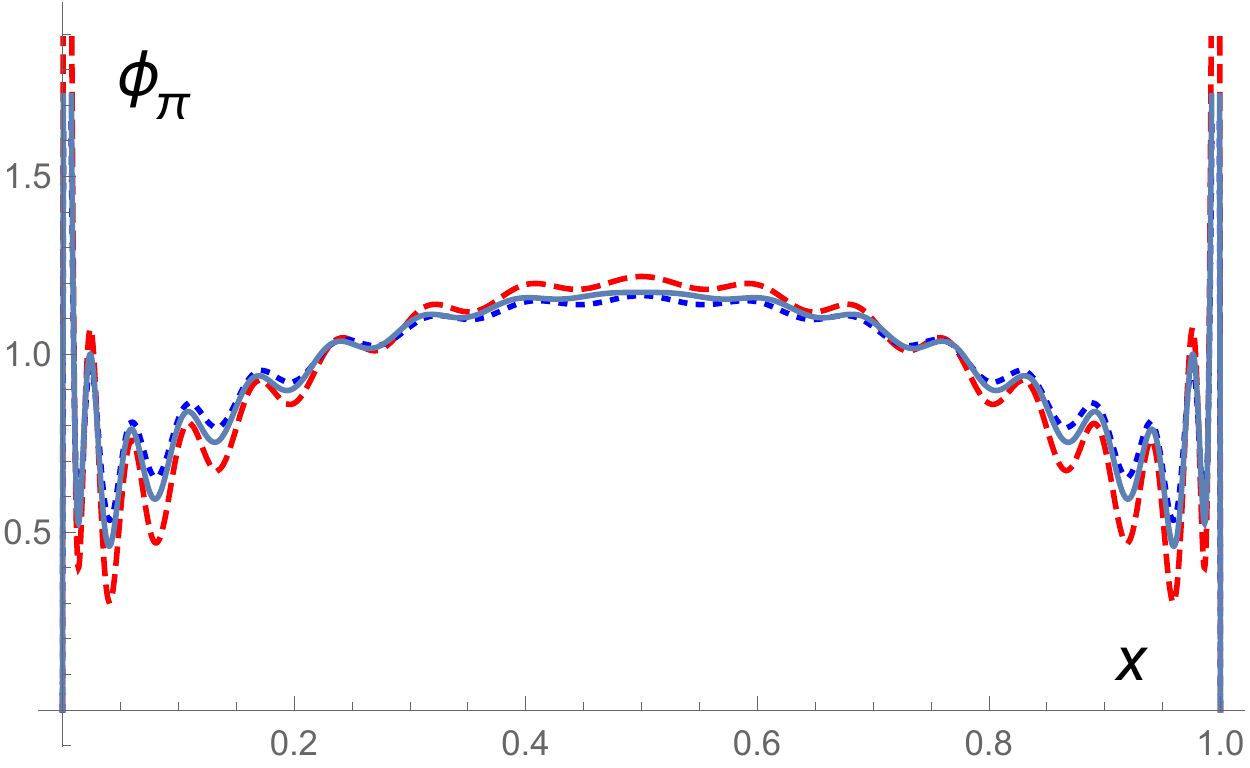}\hspace{0.5cm}
\includegraphics[scale=0.4]{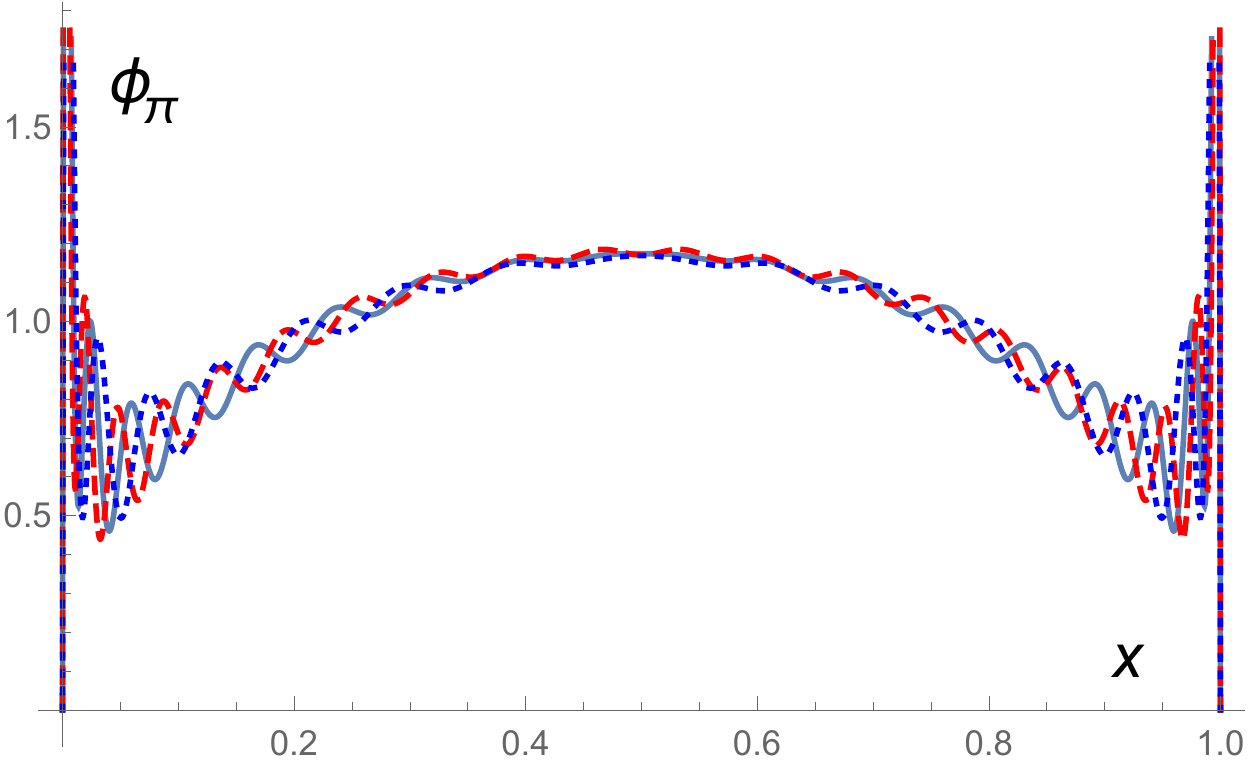}

(a) \hspace{5.0 cm} (b)
\caption{\label{fig11}
Dependencies of the pion LCDA $\phi_\pi(x)$ for (a) $\lambda=0.10$ ($\Lambda=7.45$ GeV$^2$, dashed line), 
0.20 ($\Lambda=7.36$ GeV$^2$, solid line) and $0.30$ ($\Lambda=7.31$ GeV$^2$, dotted line) with $N=18$,
and (b) $\lambda=0.20$ with $N=16$ ($\Lambda=6.35$ GeV$^2$, dotted line), 18 ($\Lambda=7.36$ GeV$^2$, solid line), 
and 20 ($\Lambda=8.39$ GeV$^2$, dashed line).}
\end{center}
\end{figure}

We have pointed out that the pion LCDA can be determined at any scale $\mu$ in principle by tuning $\mu$ in Eq.~(\ref{pn})
and choosing the corresponding condensates defined in Eq.~(\ref{mqq}). However, we have also made clear that the 
current OPE inputs are not complete; namely, the $\mu$ dependence of the inputs is not accurate strictly speaking. 
Therefore, we investigate how well our formalism is compatible with the QCD evolution: we solve for the pion LCDA 
at another scale, say, $\mu=1.5$ GeV, directly from the inputs, evolve the pion LCDA obtained previously at 
$\mu=2$ GeV to this lower $\mu$, and then compare the two results. It is found that the shapes of the pion LCDA 
in the former approach become independent of the 
regularization parameter $\lambda$ as $\lambda>0.1$. The $x$ dependencies of the pion LCDA $\phi_\pi(x)$ for $\lambda=0.10$, 
0.20 and 0.30 with the dimension $N=18$, corresponding to the transition scales $\Lambda=7.45$ GeV$^2$, $7.36$ GeV$^2$, 
and $7.31$ GeV$^2$, respectively, are displayed in Fig.~\ref{fig11}(a). The three curves overlap reasonably well, 
reflecting the stability of the solutions. Especially, the curve for $\lambda=0.20$ is smooth in the intermediate 
$x$ region. The curves show stronger oscillations near the endpoints of $x$, which might be due to the larger 
dimension-six condensate contributions at a lower $\mu$, and less perfect cancellation among various terms in the 
inverse matrix method. We have confirmed that stable solutions for the pion LCDA at the scale $\mu=2.5$ GeV
are as smooth as those in Fig.~\ref{fig8}(b). The set of Gegenbauer coefficients corresponding to $\lambda=0.20$ is given by
\begin{eqnarray}
& &(a_2^\pi,a_4^\pi,a_6^\pi, a_8^\pi, a_{10}^\pi,a_{12}^\pi,\cdots,a_{32}^\pi,a_{34}^\pi)|_{\mu=1.5\,{\rm GeV}}\nonumber\\
&=&(0.2963, 0.2661, 0.2694, 0.2700, 0.2649, 0.2550, \cdots, 0.0379, 0.0109),
\label{ga9}
\end{eqnarray}
whose elements are also all under control. We then exhibit $\phi_\pi(x)$ for $N=16$, 18, and 20 with $\lambda=0.20$, 
corresponding to $\Lambda=6.35$ GeV$^2$, 7.36 GeV$^2$, and 8.39 GeV$^2$, respectively, in Fig.~\ref{fig11}(b). The 
three curves also overlap in the intermediate $x$ region, implying that the solutions are stable with the variations
of $N$ and $\Lambda$, and oscillate strongly near the endpoints of $x$ for the similar reason.
The bands in Fig.~\ref{fig11} represent the theoretical uncertainty in our framework, which is larger than in 
the case for $\mu=2$ GeV.

\begin{figure}
\begin{center}
\includegraphics[scale=0.5]{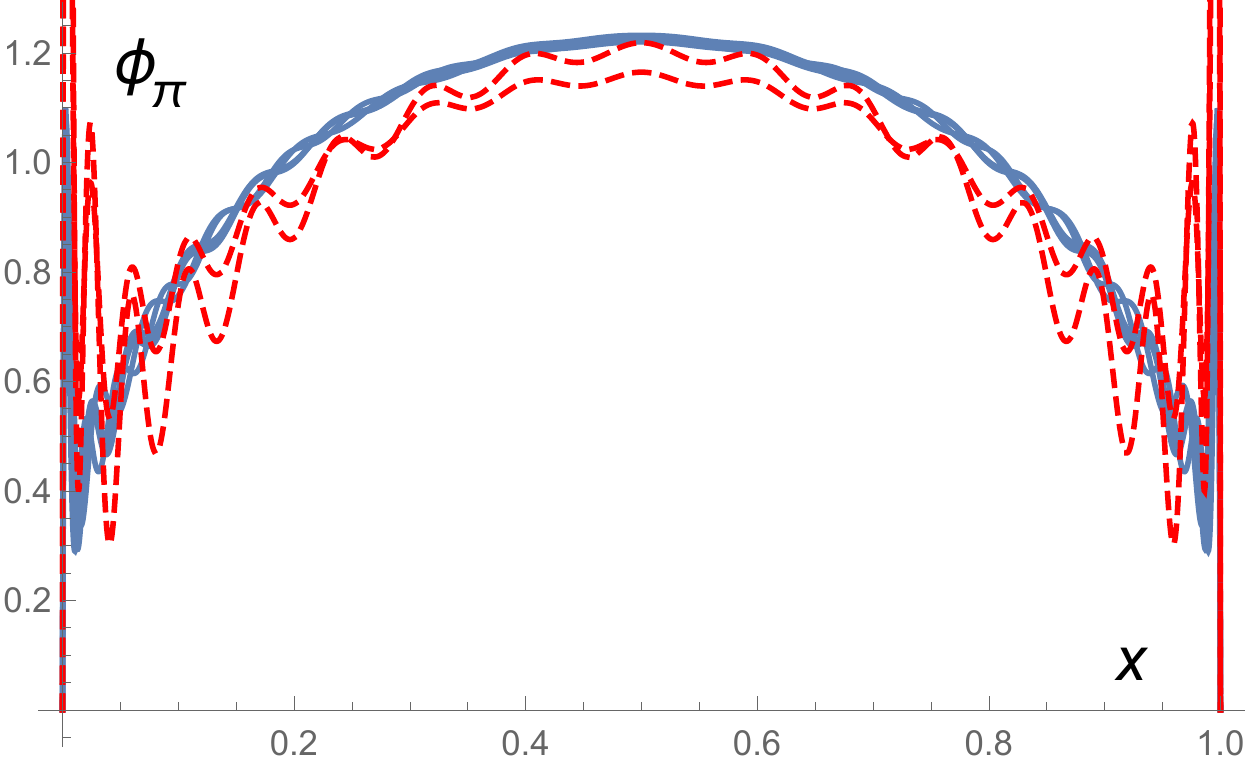}\hspace{0.5cm}

\caption{\label{fig12}
Comparison of the pion LCDA $\phi_\pi(x)$ from the evolution of those in Figs.~\ref{fig8}(b) and \ref{fig8}(c)
(solid lines), and from Fig.~\ref{fig11}(a) for $\lambda=0.10$ and $0.30$ (dashed lines).}
\end{center}
\end{figure}

It has been known that a Gegenbauer coefficient $a_n^\pi$ of the pion LCDA obeys the QCD evolution
\begin{eqnarray}
a_n^\pi(\mu)=a_n^\pi(\mu_0)E_n(\mu,\mu_0),\;\;\;\;n=2,4,6,\cdots
\end{eqnarray}
where the initial scale $\mu_0$ is set to 2 GeV, and the evolution factor is written as
\begin{eqnarray}
E_n(\mu,\mu_0)=\left[\frac{\alpha_s(\mu)}{\alpha_s(\mu_0)}\right]^{\gamma_n^{(0)}/(2\beta_0)},
\end{eqnarray}
with the leading-order anomalous dimension
\begin{eqnarray}
\gamma_n^{(0)}=8C_F\left[\psi(n+2)+\gamma_E-\frac{3}{4}-\frac{1}{2(n+1)(n+2)}\right],
\end{eqnarray}
$C_F=4/3$ being a color factor and $\gamma_E=0.5772$ being the Euler constant. We evolve the Gegenbauer
coefficients at $\mu=2$ GeV associated with the curves in Figs.~\ref{fig8}(b) and \ref{fig8}(c)
for various $\lambda$ and $N$ down to $\mu=1.5$ GeV, and construct the pion LCDAs in the Gegenbauer
expansion as shown in Fig.~\ref{fig12}. The band formed by these curves is broadened a bit by 
the evolution effect. The curves from solving the dispersion relations directly for $\lambda=0.1$ and 
$\lambda=0.3$ in Fig.~\ref{fig11}(b), selected as the representative ones, are then shown for comparison. 
It is observed that the former is slightly above the latter in the intermediate $x$ region, 
but they overlap when the theoretical uncertainties are taken into account. This rough agreement supports 
that our formalism is compatible with the QCD evolution within theoretical uncertainties.

\section{CONCLUSION}

We have handled dispersion relations obeyed by a nonperturbative correlation function 
in a novel way, much different from that of conventional sum rules. It follows our earlier proposal for
solving dispersion relations as an inverse problem with the OPE of the correlation function as inputs.
This formalism does not assume the quark-hadron duality for the continuum contribution,
does not involve a continuum threshold in the parametrization of a spectral density, requires no 
Borel transformation to suppress the continuum contribution and higher power corrections, 
and needs no discretionary stability criteria on the balance between perturbative and condensate 
contributions. With these merits, extracting all the moments of the leading-twist 
pion LCDA becomes possible as demonstrated in this work. The idea is to expand the
continuum function in an orthogonal polynomial basis, which is formed by the generalized
Laguerre polynomials, and to solve for the unknown coefficients in the expansion, together with
the moments appearing in the pion pole contribution, in the inverse matrix method. We have pointed put 
that the power-suppressed logarithm $\ln(-q^2)/(q^2)^3$ in the OPE must be reformulated  into a power 
series in $1/q^2$ by means of a dispersive integral, before the inverse matrix method can be implemented. 
It has been shown that solutions for these unknowns are stable with respect to the number $N$ of polynomials 
in the expansion, and to the variation of the transition scale $\Lambda$, which is introduced through the 
ultraviolet regularization for the spectral density. Inputting the quark and gluon condensates 
in the literature, we have obtained the moments of the pion LCDA at the scale $\mu=2$ GeV close to those from other 
approaches. 

We have emphasized that it is highly nontrivial to acquire the $x$ dependence of the pion LCDA, even when 
all the moments are available, because the conversion from the moments to the Gegenbaer coefficients is also
a challenging ill-posed problem. Therefore, we have further extended our framework to the analysis of the
dispersion relations for the Gegenbauer coefficients, which come from the linear combination of those 
for the moments. To smear the strong fluctuation in solutions caused by the ill-posed nature, a 
regularization has been introduced into the inverse matrix method, whose strength is
characterized by a parameter $\lambda$. It has been observed that solutions for the
Gegenbauer coefficients with excellent convergence exist, which are insensitive to the variations of $\lambda$, $N$
and $\Lambda$ in finite ranges. Furthermore, the pion LCDA from summing the contributions up to 18 
Gegenbauer polynomials reveals a smooth shape, in agreement with that from the 
dynamical-chiral-symmetry-breaking-improved kernel for the Dyson-Schwinger equations, and similar to that from
the recent lattice QCD derivation based on the quasi-light-front correlation function, but different from
that governed by a finite number of moments from conventional sum rules. We have verified that the 
asymptotic form is retrieved for the pion LCDA in the absence of the condensates, and that our 
formalism is compatible with the QCD evolution: the solution for the pion LCDA with the condensate inputs 
at a different scale $\mu=1.5$ GeV matches the one obtained by evolving the Gegenbauer coefficients 
from $\mu=2$ GeV to this lower scale within theoretical uncertainties.

We have surveyed the various sources of theoretical uncertainties in our calculations, which are 
summarized below. The uncertainties from the condensates $\langle m_q\bar q q\rangle$ and 
$m_q \langle g_s\bar{q}\sigma TGq\rangle$  in the OPE inputs are negligible. The variations of the 
dimension-four condensate $\langle\alpha_s G^2\rangle$ and the dimension-six condensates dominate the 
uncertainties, amounting up to order of $10\%$. The uncertainty in our method, resulting from the 
finite stability intervals of the parameters $\lambda$, $N$ and $\Lambda$, is about 2\%-3\%, as elaborated
via the evaluations of the moment $\langle\xi^2\rangle$ and the Gegenbauer coefficient $a_2^\pi$. 
The choices of the subtraction terms for the continuum function, i.e., of the ultraviolet regularization 
for the spectral density causes 1\% error at most. Overall speaking, it is reasonable to claim order of 10\%
theoretical uncertainties in our solutions for the pion LCDA. It should be remembered that the zeroth moment
$\langle\xi^0\rangle$, related to the normalization of the pion LCDA, is not equal to unity, and 
depends on the scale $\mu$ under the current incomplete OPE inputs. We have employed the alternative 
interpretation of the dispersion relation for $\langle\xi^0\rangle$ as the one for the pion decay
constant on the premise $\langle\xi^0\rangle=1$. By considering Eqs.~(\ref{so}) and (\ref{soa}), the resultant 
$\mu$-dependent pion decay constant cancels in the ratios, and the issue about the $\mu$ dependence of 
$\langle\xi^0\rangle$ is resolved. The uncertainty associated with the above treatment was not taken 
into account in the present work. We urge an inclusion of higher-power contributions, such as that from 
dimension-eight condensates, into the OPE for the relevant correlation function, which are expected to 
rectify the normalization of the pion LCDA efficiently.

We have demonstrated how to extract information on the leading-twist pion LCDA as much as possible
from the known dispersion relations. The framework developed here is ready for applications to studies 
of other hadron LCDAs, and likely to be extended to determinations of parton distribution 
functions for inclusive processes. It goes beyond analyses usually limited to a finite number of moments in 
the literature, and serves as a simple analytical approach to nonperturbative observables. The precision of 
predictions from this formalism can be improved by adding higher-order and higher-power contributions to 
the OPE systematically, and by fixing the values of the quark and gluon condensates, which are supposed to 
be universal inputs.

\section*{Acknowledgement}

We thank N. Stefanis, X.G. Wu and T. Zhong for helpful discussions. This work was supported in part 
by the Ministry of Science and Technology of the Republic of China. under Grant No.
MOST-110-2811-M-001-540-MY3.

\appendix

\section{REFORMULATION OF POWER-SUPPRESSED LOGARITHM}

We provide the details of reformulating the power-suppressed logarithm $\ln(-q^2/\mu^2)/(q^2)^3$ into
a power series in $1/q^2$ by means of a dispersive integral. We first work on a simple power-suppressed 
logarithm $L^{(1)}(q^2)=\ln(-q^2/\mu^2)/q^2$, and produce it by a power series in $1/q^2$. The insertion of
$L^{(1)}(q^2)$ into the contour integral in Eq.~(\ref{10}) leads to
\begin{eqnarray}
L^{(1)}(q^2)
=\frac{1}{2\pi i}\int_{C_r} ds\frac{L^{(1)}(s)}{s-q^2}+
\frac{1}{\pi}\int_r^R ds\frac{{\rm Im}L^{(1)}(s)}{s-q^2}
+\frac{1}{2\pi i}\int_{C_R} ds\frac{L^{(1)}(s)}{s-q^2}.\label{4}
\end{eqnarray}
We apply the variable change $s=r\exp(i\theta)$ to the first integral on the right-hand side, and expand 
the denominator $r\exp(i\theta)-q^2$ up to the leading power in $r\exp(i\theta)/q^2$, since the 
next-to-leading power vanishes in the $r\to 0$ limit. The argument of the logarithm in the numerator, 
$-r\cos\theta-ir\sin\theta$, rotates from the third quadrant counterclockwise as $\theta$ increases from 
zero. It implies that the minus sign can be set to $\exp(-i\pi)$ and the argument becomes 
$-r\exp(i\theta)=r\exp[i(\theta-\pi)]$. This assignment guarantees that the first integral gives a real value:
\begin{eqnarray}
\int_0^{2\pi} d\theta\ln\frac{-r\exp(i\theta)}{\mu^2}
=\int_0^{2\pi} d\theta\ln\frac{r}{\mu^2} +
i\int_0^{2\pi} d\theta(\theta-\pi)=2\pi\ln\frac{r}{\mu^2}.
\end{eqnarray}

The substitution of ${\rm Im}L^{(1)}(s)=-\pi/s$ into the second term on the right-hand side 
of Eq.~(\ref{4}) yields
\begin{eqnarray}
L^{(1)}(q^2)&=&\frac{1}{2\pi q^2}\int_0^{2\pi} d\theta\ln\frac{r\exp[i(\theta-\pi)]}{\mu^2}
-\int_r^R \frac{ds}{s(s-q^2)}
+\frac{1}{2\pi i}\int_{C_R} ds\frac{L^{(1)}(s)}{s-q^2},\label{5}
\end{eqnarray} 
which reduces to
\begin{eqnarray}
L^{(1)}(q^2)&=&\frac{1}{q^2}\ln\frac{r}{\mu^2}
+\frac{1}{q^2}\left(\ln\frac{r-q^2}{r}-\ln\frac{R-q^2}{R}\right)
+\frac{1}{2\pi i}\int_{C_R} ds\frac{L^{(1)}(s)}{s-q^2}.
\end{eqnarray}
The above expression verifies that the $r\to 0$ limit can be taken safely, and $L^{(1)}(q^2)=\ln(-q^2/\mu^2)/q^2$
has been reproduced, as the term $\ln[(R-q^2)/R]$ and the last integral are dropped at large $R$. 
We then arrive at
\begin{eqnarray}
L^{(1)}(q^2)&=&\lim_{r\to 0}\frac{-rL^{(1)}(-r)}{q^2}+
\lim_{r\to 0}\frac{1}{\pi}\int_r^R ds\frac{{\rm Im}L^{(1)}(s)}{s-q^2}
+\frac{1}{2\pi i}\int_{C_R} ds\frac{L^{(1)}(s)}{s-q^2},\label{pr}
\end{eqnarray}
where the first term is of power $1/q^2$, and the second term can be cast into a power series 
in $1/q^2$ by expanding the denominator of the integrand.

We then extend the above procedure to $L(q^2)=\ln(-q^2/\mu^2)/(q^2)^3$, starting with the identity
\begin{eqnarray}
L(q^2)&=&\frac{1}{2\pi}\int_0^{2\pi} \frac{d\theta\exp(-2i\theta)}{r^2[q^2-r\exp(i\theta)]}
\ln\frac{r\exp[i(\theta-\pi)]}{\mu^2}-\int_r^R \frac{ds}{s^3(s-q^2)}
+\frac{1}{2\pi i}\int_{C_R} ds\frac{L(s)}{s-q^2}.\label{6}
\end{eqnarray}
The denominator $q^2-r\exp(i\theta)$ in the first integral on the right-hand side is expanded up to the second power in 
$r\exp(i\theta)/q^2$, and three pieces survive in the $r\to 0$ limit:
\begin{eqnarray}
\frac{1}{2\pi q^2r^2}\int_0^{2\pi} d\theta\exp(-2i\theta)
\ln\frac{-r\exp(i\theta)}{\mu^2}&=&
\frac{1}{2\pi q^2r^2}\int_0^{2\pi} d\theta\exp(-2i\theta)i\theta=-\frac{1}{2q^2r^2},\nonumber\\
\frac{1}{2\pi(q^2)^2r}\int_0^{2\pi} d\theta\exp(-i\theta)
\ln\frac{-r\exp(i\theta)}{\mu^2}&=&
\frac{1}{2\pi(q^2)^2r}\int_0^{2\pi} d\theta\exp(-i\theta)i\theta=-\frac{1}{(q^2)^2r},\nonumber\\
\frac{1}{2\pi(q^2)^3}\int_0^{2\pi} d\theta
\ln\frac{-r\exp(i\theta)}{\mu^2}&=&\frac{1}{(q^2)^3}\ln\frac{r}{\mu^2}.\label{7}
\end{eqnarray}
The second integral on the right-hand side of Eq.~(\ref{6}) gives
\begin{eqnarray}
-\int_r^R \frac{ds}{s^3(s-q^2)}=\frac{1}{(q^2)^3}\ln\frac{r-q^2}{r}+\frac{1}{(q^2)^2r}
+\frac{1}{2q^2r^2}-\frac{1}{(q^2)^3}\ln\frac{R-q^2}{R}-\frac{1}{(q^2)^2R}
-\frac{1}{2q^2R^2}.
\end{eqnarray}
It is immediately seen that the three pieces in Eq.~(\ref{7}) cancel the first three terms in the above expression.
The similar steps then lead Eq.~(\ref{6}) to Eq.~(\ref{i3}).

\end{document}